\documentclass[journal]{rmaa}

\usepackage{amsmath}
\usepackage{paralist}
\usepackage{psfrag,color}
\usepackage[latin1]{inputenc}
\usepackage{ulem} 

\title{Periodicity detection in AGN with the boosted tree method}

\author{
 S. B. Soltau\altaffilmark{1,2} 
 and
 L. C. L. Botti\altaffilmark{2,3}
}

\altaffiltext{1}{Federal University of Alfenas, Brazil.}
\altaffiltext{2}{Center for Radio Astronomy and Astrophysics Mackenzie, S\~ao Paulo, Brazil.}
\altaffiltext{3}{Brazilian National Institute for Space Research, S\~ao Jos\'e dos Campos, Brazil.}

\shortauthor{Soltau \& Botti}

\shorttitle{Periodicity Detection in AGN}

\fulladdresses{
\item Samuel Bueno Soltau: Department of Physics, Institute of Exact Sciences.
 Federal University of Alfenas, Minas Gerais, Brazil.
 (samuel.soltau@unifal-mg.edu.br).
 orcid 0000-0002-7211-2533

 \item Luiz Claudio Lima Botti: Center for Radio Astronomy and Astrophysics Mackenzie, Engineering School,
 Mackenzie Presbyterian University, S\~ao Paulo, Brazil.
 and Astrophysics Division, Brazilian National Institute for Space Research,
 S\~ao Jos\'e dos Campos, S\~ao Paulo, Brazil
 (luizquas@yahoo.com.br).
 orcid 0000-0003-1424-0796}

\listofauthors{S. B. Soltau \& L. C. L. Botti}

\indexauthor{Soltau, S. B.}
\indexauthor{Botti, L. C. L.}

\abstract{We apply a machine learning algorithm called XGBoost to explore the periodicity of two radio sources: PKS~1921-293 (OV~236) and PKS~2200+420 (BL~Lac), both radio frequency dataset obtained from University of Michigan Radio Astronomy Observatory (UMRAO), at $4.8~\mathrm{GH\lowercase{z}}$, $8.0~\mathrm{GH\lowercase{z}}$, and $14.5~\mathrm{GH\lowercase{z}}$, between 1969 to 2012.
From this methods, we find that the XGBoost provides the opportunity to use a machine learning based methodology on radio dataset and to extract information with strategies quite different from those traditionally used to treat time series and to obtain periodicity through the classification of recurrent events.
The results were compared with other methods from others works that examined the same dataset and exhibit good agreement with them.}

\resumen{Aplicamos un algoritmo de aprendizaje automático llamado XGBoost para explorar la periodicidad de dos fuentes de radio: PKS~1921-293 (OV~236) y PKS~2200+420 (BL~Lac), ambos conjuntos de datos de radiofrecuencia obtenidos de Radio Astronomía de la Universidad de Michigan Observatorio (UMRAO), a $4.8~\mathrm{GH\lowercase{z}}$, $8.0~\mathrm{GH\lowercase{z}}$, y $14.5~\mathrm{GH\lowercase{z}}$, entre 1969 y 2012.
A partir de estos métodos, encontramos que XGBoost brinda la oportunidad de utilizar una metodología basada en aprendizaje automático en el conjunto de datos de radio y extraer información con estrategias bastante diferentes de las utilizadas tradicionalmente para tratar series temporales y obtener periodicidad a través de la clasificación de eventos recurrentes.
Los resultados se compararon con los obtenidos en otros trabajos que examinaron el mismo conjunto de datos y muestraron resultados compatibles.}

\addkeyword{galaxies: active}
\addkeyword{galaxies: BL Lacertae objects: general}
\addkeyword{galaxies: quasars: general}
\addkeyword{methods: data analysis}
\addkeyword{methods: numerical}

\begin{document}
\maketitle

\section{Introduction}
\label{sec:intro}

Ever since discovery of first radio sources~\citep{MatthewsSandage1963,Schmidt1963} in 1963, 
a considerable amount of workforce and computing resources have been invested in exploring the observable Universe to detect radio sources. Quasar and \textit{BL~Lacertae} are supermassive rotating black hole, with jet ejection and rotation axes, which emit in radio, X-rays and gamma energies. Their radio signals are observable when the axis of their emission cone is directed along the line of sight to the instrument. Subsequently, they have also been observed throughout the electromagnetic spectrum. An updated review of the various observational properties of quasars and other kinds of active galaxy nuclei (AGN) can be found in~\citet{VeronCetty2010}.

AGNs, particularly quasars, have been studied at many radio frequencies to understand the mechanisms and regimes of energies involved in the phenomenon. As a result, a unified model was elaborated in which the different denominations given to the AGN are derived from the orientation of the jets in relation to the viewing angle of the observer~\citep{Antonucci1993, UrryPadovani1995, BeckmannShrader2012}.

The variability is the aspect of AGN that attracts the most attention. Some radio sources have periodicity measured in the scale of years, but due to the delay between the measurements made in the several frequencies, it is difficult to accurately specify periodicity. Delay makes it difficult to study time series when comparing light curves at different frequencies. In addition, the data set comprises a time series of irregular sampling due to various factors influencing the acquisition of astronomical data from ground stations, such as weather conditions, system maintenance, receivers, etc. These sampling difficulties produce unequally spaced time series, which impose limitations on more conventional methods of analysis.

Multifrequency studies explore distinct aspects of compact radio sources, in particular, flux density variations, to determine periodicities in light curves \citep{AbrahamKaufmannBotti1982, Botti1983, BottiAbraham1987, BottiAbraham1988, Botti1990, BottiIAU1994, AllerAllerHughes2009, AllerAller2010, AllerAller2011}.
Methods for determining periodicities in the radio range include Fourier Transform, Lomb-Scargle Periodogram, Wavelet Transform and Cross Entropy, among others \citep{Cincotta1995, Tornikoski1996, Auta2007, Soldi2008, Vitoriano2018}. Combination of methods, like decision trees, random forests and autoregressive models for the specific goal of exoplanets detection in stellar light curves \citep{Caceres2019}.

Advances in Artificial Intelligence have provided machine learning algorithms, such as Neural Networks, Ensemble and Deep Learning \citep{LeCunBengioHinton2015}, that have provided astrophysical studies and provided computational approaches dissimilar to previous methods, including potential applications for radio source analyses~\citep{Witten2016}.

Motivated by the successful performance of XGBoost~\citep{Chen2016} in international challenges on Machine Learning~\citep{Kaggle2018} and animated by the many different kinds of results presented by~\citet{Pashchenko2017}, \citet{Smirnov2017}, \citet{Bethapudi2018}, \citet{AbayBoyce2018}, \citet{Roestel2018}, \citet{Saha2018}, \citet{LamKipping2018}, \citet{ShuKoposov2019}, \citet{LiuRyley2019}, \citet{Askar2019}, \citet{Calderon2019}, \citet{ChongYang2019}, \citet{JinZhang2019}, \citet{Li2019}, \citet{Menou2019}, \citet{PlavinKovalev2019}, \citet{WangPan2019}, \citet{YiZesheng2019}, \citet{Li2020}, \citet{LinLiLuo2020}, \citet{Hinkel2020}, \citet{Tamayo2020} and~\citet{Tsizh2020} we decided to try out how this kind of algorithm would perform specific tasks related to the treatment of time series in radio datasets of AGNs, such as in light curves of quasars and BL Lacs. For this reason we selected two well-studied objects by the astrophysical community, the PKS~1921-293 (OV~236) quasar and PKS~2200+420 (BL~Lac) for case study.

The outline of this paper is as follows. 
In \S~\ref{sec:datasets}, the AGNs datasets from the UMRAO survey along with the features used for training and tests are described. 
In \S~\ref{sec:xgboost}, we show the machine learning algorithms applied to the AGNs datasets. The implementation of XGBoost are described in detail.
We discuss the feature selection procedure methods in \S~\ref{sec:method}.
We report and discuss the results of machine learning algorithms for the selected tasks in \S~\ref{sec:results}.
We then summary and conclude in \S~\ref{sec:summconc}.

\section{Instrument and Datasets}
\label{sec:datasets}

The Michigan Radio Astronomy Observatory (UMRAO), has a parabolic reflector antenna of about $26$ meters in diameter. This radio telescope has been used extensively since 1965 to monitor continuous full-flux density and linear polarization of variable extragalactic radio source in the frequencies of $4.8~\mathrm{GH\lowercase{z}}$ ($6.24~\mathrm{cm}$), $8.0~\mathrm{GH\lowercase{z}}$ ($3.75~\mathrm{cm}$) and $14.5~\mathrm{GH\lowercase{z}}$ ($2.07~\mathrm{cm}$). More details about UMRAO characteristics and their astrophysics applications are reviewed and can be found in~\citet{Aller1992, Aller2017}.

In our study, we used UMRAO datasets for the PKS~1921-293 (OV~236) and PKS~2200+420 (BL~Lac), in the time intervals presented in Table~\ref{tab:databases}.

\begin{table}[!ht]\centering
  \setlength{\tabnotewidth}{0.5\linewidth}
  \caption{Irregularly spaced time series. The UMRAO datasets acquired in frequencies of $4.8~\mathrm{GH\lowercase{z}}$, $8.0~\mathrm{GH\lowercase{z}}$ and $14.5~\mathrm{GH\lowercase{z}}$ from radio sources PKS~1921-293 (OV~236) and PKS~2200+420 (BL~Lacertae).}
  \label{tab:databases}
  \begin{tabular}{r c c}
   \toprule
   Frequency & PKS~2200+420 & PKS~1921-293 \\
   \midrule             
   $4.8~\mathrm{GH\lowercase{z}}$ & $1977$--$2012$ & $1979$--$2011$ \\
   $8.0~\mathrm{GH\lowercase{z}}$ & $1968$--$2012$ & $1974$--$2011$ \\
   $14.5~\mathrm{GH\lowercase{z}}$ & $1974$--$2012$  & $1975$--$2011$ \\
   \bottomrule
  \end{tabular}            
\end{table}

It is worth mentioning that these are irregularly spaced time series, so the years indicated in the Table~\ref{tab:databases} refer to the range of years covered in this study. The differences in the years of radio datasets start for each frequency, for both objects of study, is due to the fact that the UMRAO began to operate in each one of the frequencies in different epochs.

For all objects in dataset collection and at all operating frequencies, UMRAO provides daily time series\footnote{Publicly available in \url{https://dept.astro.lsa.umich.edu/datasets/umrao.php} with permission.}. Due to several inherent aspects of observation, such as weather and instrumental maintenance, the data sets are ir\-regularly spaced, requiring treatment before being used in the research. The procedures adopted during processing for this purpose will be described later in \S~\ref{sec:method}.

The Figs.~\ref{fig:rawdatasetsbllac} and~\ref{fig:rawdatasetsov236} in Appendix~\ref{sec:appRawData}, show the light curves for the objects PKS~2200+420 (BL~Lac) and PKS~1921-293 (OV~236), respectively, arranged in graphs according to the same time interval for comparison purposes.

The characteristics of the radio source PKS~2200+420 (BL~Lac)\footnote{Available data \url{http://simbad.u-strasbg.fr/simbad/sim-id?Ident=NAME\%20BL\%20Lac}.} used in this study are galactic coordinates $92.5896$ $-10.4412$, equatorial coordinates (J2000) RA~$22$h$02$m$43,291$s DE~$+42^{o}16'39,98$, constelation \textit{Lacerta}, apparent magnitude $V=14.72$, absolute magnitude $MV=~-22.4$ and redshift $z~\approx~0.069$.

The characteristics of the radio source PKS~1921-293 (OV~236)\footnote{Available data \url{http://simbad.u-strasbg.fr/simbad/sim-id?Ident=PKS\%201921-293}.} are equatorial coordinates (J2000) RA~$19$h$24$m$51.056$s DE~$-29^{o}14'30,11$, galactic coordinates $9.3441$ $-19.6068$, constelation \textit{Sagittarius}, apparent magnitude $V=17.5$, absolute magnitude $MV=-24.6$ and redshift $z \approx 0.353$.

The UMRAO datasets are provided in digital files in American Standard Code for Information Interchange (ASCII) coding standard and contain, listed in daily sequences, the acquisition date in Modified Julian Date format, the flux density and the associated measurement error, both in jansky. Table~\ref{tab:obslines} shows the number of observations per frequency for each radio source addressed in this study.
\begin{table}[h]\centering
\setlength{\tabnotewidth}{0.5\linewidth}
\caption{Observations in time series. The UMRAO datasets acquired in frequencies of $4.8~\mathrm{GH\lowercase{z}}$, $8.0\mathrm{GH\lowercase{z}}$ and $14.5\mathrm{GH\lowercase{z}}$ from radio sources PKS~1921-293 (OV~236) and PKS~2200+420 (BL~Lac).}
\label{tab:obslines}
\begin{tabular}{r c c}
\toprule
Frequency & PKS~2200+420 & PKS~1921-293 \\
\midrule
  $4.8~\mathrm{GH\lowercase{z}}$ & $692$ & $~\,\,618$ \\
  $8.0~\mathrm{GH\lowercase{z}}$ & $843$ & $~\,\,910$ \\
$14.5~\mathrm{GH\lowercase{z}}$ & $962$ & $1\,035$ \\
\bottomrule
\end{tabular} 
\end{table}

\section{XGBoost}
\label{sec:xgboost}

XGBoost, an acronym for eXtreme Gradient Boosting, is a set of machine learning methods boosted tree based, packaged in a library designed and optimized for the creation of high performance algorithms~\citep{Chen2016}. Its popularity in the machine learning community has grown since its inception in 2016. This model was also the winner of High Energy Physics meets Machine Learning Kaggle Challenge~\citep{Kaggle2018}. In astrophysics, XGBoost was recently used for the classification of pulsar signals from noise~\citep{Bethapudi2018} and also to search for exoplanets extracted from the PHL-EC (Exoplanet Catalog hosted by the Planetary Habitability Laboratory)\footnote{The latest updated (July 2, 2018) dataset can be downloaded from the PHL website: \url{http://phl.upr.edu/projects/habitable-exoplanets-catalog/data/database}.} using physically motivated features with the help of supervised learning~\citep{Saha2018}.

Gradient boosting is a technique for building models in machine learning. The idea of boosting originated in a branch of machine learning research known as computational learning theory. There are many variants on the idea of boosting~\citep{Witten2016}. The central idea of boosting came out of the idea of whether a ``weak learner'' can be modified to become better. The first realization of boosting that saw great success in application was Adaptive Boosting or AdaBoost and is designed specifically for classification. The weak learners in AdaBoost are decision trees with a single split, called decision stumps for their shortness~\citep{Witten2016}.

AdaBoost and related algorithms were recast in a statistical framework and became known as Gradient Boosting Machines. The statistical framework cast boosting as a numerical optimization problem where the objective is to minimize the loss function of the model by adding weak learners using a gradient descent like procedure. The Gradient Boosting algorithm involves three elements:
\begin{inparaenum}[(i)]
\item A loss function to be optimized such as cross entropy for classification or mean squared error for regression problems.
\item A weak learner to make decisions, integrating a decision tree.
\item An additive model, used to add weak learners to minimize the loss function. New weak learners are added to the model in an effort to correct the residual errors of all previous trees. The result is a powerful modeling algorithm.
\end{inparaenum}

XGBoost works in the same way as Gradient Boosting but with the addition of Adaboost-like feature of assigning weights to each sample. In addition to supporting all key variations of the technique, the real interest is the speed provided by the implementation, including:
\begin{inparaenum}[(i)]
\item Parallelization of tree construction using all computer CPU cores during training.
\item Distributed computing for training very large models using a cluster of computers.
\item Out-of-core Computing for very large datasets that don't fit into memory.
\item Cache optimization of data structures and algorithms to make best use of hardware~\citep{MitchellFrank2017}.
\end{inparaenum}

XGBoost core engine can parallelize all members of the ensemble (tree), give substantial speed boost and reducing computational time. In the other hand, the statistical machine-learning classification method used for supervised learning problems, where the training data with multiple features are used to forecast a target variable and the regularization techniques, are used to control over-fitting. The XGBoost method uses a non-metric classifier, and is a fairly recent addition to the suite of machine learning algorithms~\citep{Chen2016}. Non-metric classifiers are applied in scenarios where there are no definitive notions of similarity between feature vectors.

Traditionally, gradient boosting implementations are slow because of the sequential nature in which each tree must be constructed and added to the model. XGBoost solve the slowness problem put trees to work toghether, creating the concept of forest. This approach improve the performance in the development of XGBoost has resulted in one of the best modeling algorithms that can now harness the full capability of very large disponibility hardware platform (cf. benchmark tests in~\citep{ZhangQianMaoHuangHuangSi2018, HuangYen2019}).

In a typical machine-learning problem, the processes input data try to combines a large number of regression trees with a small learning rate to produce a model as output. In this case, learning means recognizing complex patterns and making intelligent decisions based on input dataset features provides by the human supervisor.

The algorithm comes up with its own prediction rule, based on which a previously unobserved sample will be classified as of a certain type, e. g. high and low activity period for give a pertinent example, with a reasonable accuracy. In order to appropriately apply a method (including preprocessing and classification), a thorough study of the nature of the data should be done; this includes understanding the number of samples in each class, the separability of the data, etc. Depending on the nature of the data, appropriate preprocessing and post processing methods should be determined along with the right kind of classifier for the task (e.g. binary classification or multiclass classification)~\citep{LeCunBengioHinton2015}.

XGBoost have distinct machines learning capabilities only: regression and classification trees. All tasks and problem to be solved, need be reduced to this two categories.
Regression trees are used for continuous dependent variables. Classification trees for categorical dependent variable. In regression tree, the value obtained by terminal nodes in the training data is the mean response of observation falling in that region. In classification tree, the value obtained by terminal node in the training data is the mode of observations falling in that region.
In this research, the developed method uses both of capabilities, as will be shown in \S~\ref{sec:method}.

XGBoost is readily available as a Python API (Application Program Interface), which is used in this work.\footnote{The source code was available at \url{https://github.com/sbs-PhD/astroph}.}

To the best of our knowledge, among XGBoost algorithm, have never been used before in AGN for regularization of time series or during the post-processing of outbursts selection candidates.

\section{Method}
\label{sec:method}

We prepare the machine to learn the features associated with the training and test data to fill irregularly spaced time series and identify the occurency of outbursts in radio sources datasets from UMRAO through machine classification algorithm XGBoost. The goal, as stated earlier, is to test the ability of the algorithm to be used in astrophysics studies of radio sources AGN-like with a reasonably high accuracy, thereby establishing the utility of this method where different approaches are useless. The classification of outburst was done with classification tree, whilst the regularization of time series was done with regression tree.

The entire method can be summarized in the following steps:
\begin{inparaenum}[(i)]
\item obtaining and preparing the data (pre-processing);
\item regularizing the time series;
\item detection of outbursts; and,
\item calculation of periodicity within defined limits of accuracy. 
\end{inparaenum}
Here we will highlight the regularization of time series and the detection of outbursts, mainly. 

\subsection{Preprocessing}
\label{sec:prepro}

Preprocessing is an essential preliminary step in any machine learning technique, as the quality and effectiveness of the following steps depend on it~\citep{Brighton2002}. 
This covers from obtaining the original UMRAO files, in ASCII digital format, to preparing the data with the application of algorithms whose purpose is to check data consistency, eliminate incomplete lines or other inconsistencies typical of experimental datasets stored in files in the format such as spurious characters, formatting, etc. All the procedures applied in this phase act direct and only on the original data, but without altering them in their fundamental characteristics.
The procedure also has the purpose of removing the beginning or end of the data in case of big time lag to the next data reducing the error propagation and the computational time will be reduced.

The original data files of the UMRAO contain all three frequencies acquired, $4.8~\mathrm{GH\lowercase{z}}$, $8.0~\mathrm{GH\lowercase{z}}$ and $14.5~\mathrm{GH\lowercase{z}}$ unsorted in the file lines. Each line corresponds to a daily measurement in a given frequency. During preprocessing, rows of the same frequency were collected and stored together in a separate file. In this way, each frequency can be treated independently for each object studied.

\begin{figure}[!h]\centering
\includegraphics[width=1.0\linewidth]{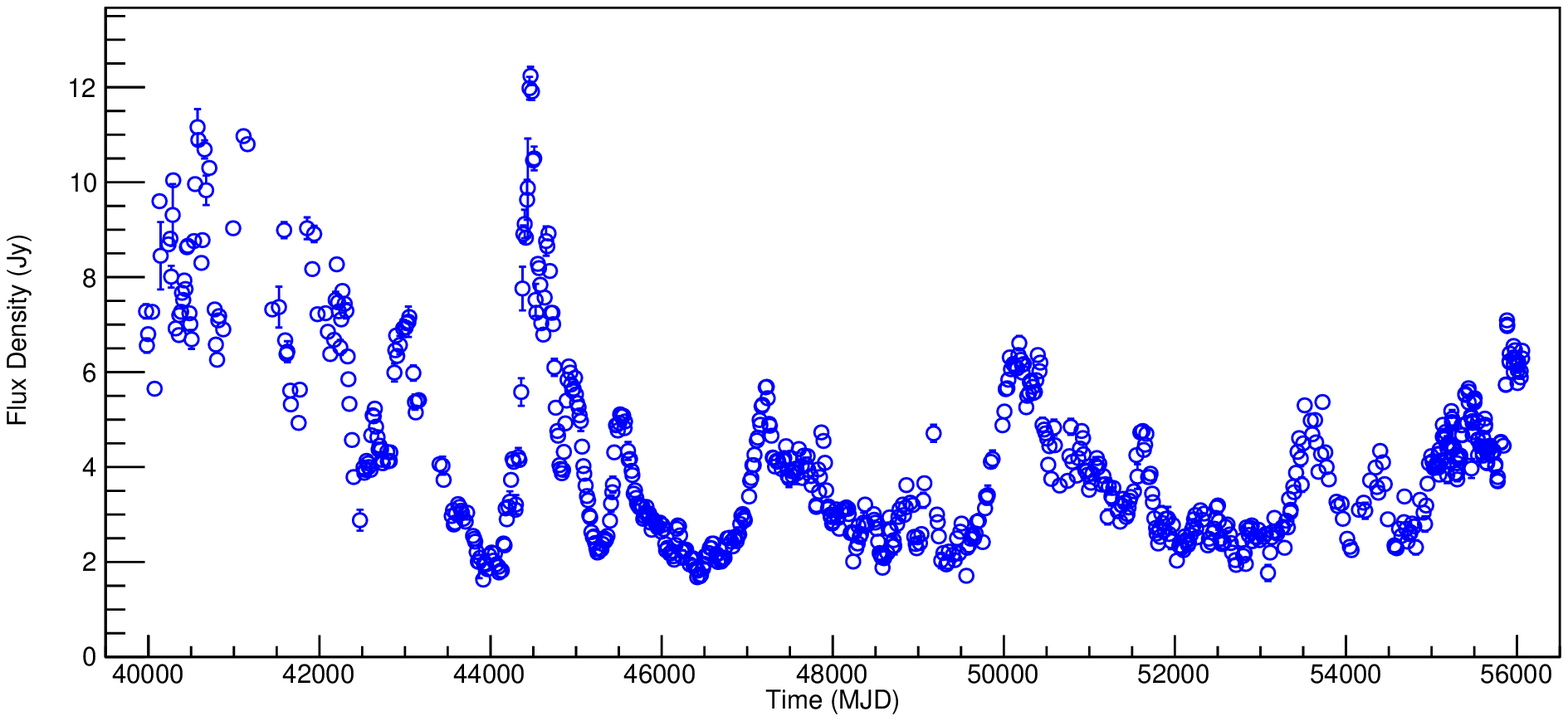}%
\caption{Light curve of PKS~2200+420 radio source, in $8.0~\mathrm{GH\lowercase{z}}$, shaw the raw dataset.}
\label{fig:rawbllac080}
\end{figure}

At the end of this step, the graphs of the original daily flux density data were plotted. For simplicity, only the $8.0~\mathrm{GH\lowercase{z}}$ is show in Fig.~\ref{fig:rawbllac080} for PKS~2200+420 (BL~Lac) radio source. All others frequencies, $4.8~\mathrm{GH\lowercase{z}}$ and $14.5~\mathrm{GH\lowercase{z}}$ are shown in Figs.~\ref{fig:rawbllac048} and~\ref{fig:rawbllac145} Appendix~\ref{sec:appPreprocess}.

\begin{figure}[!h]\centering
\includegraphics[width=1.0\linewidth]{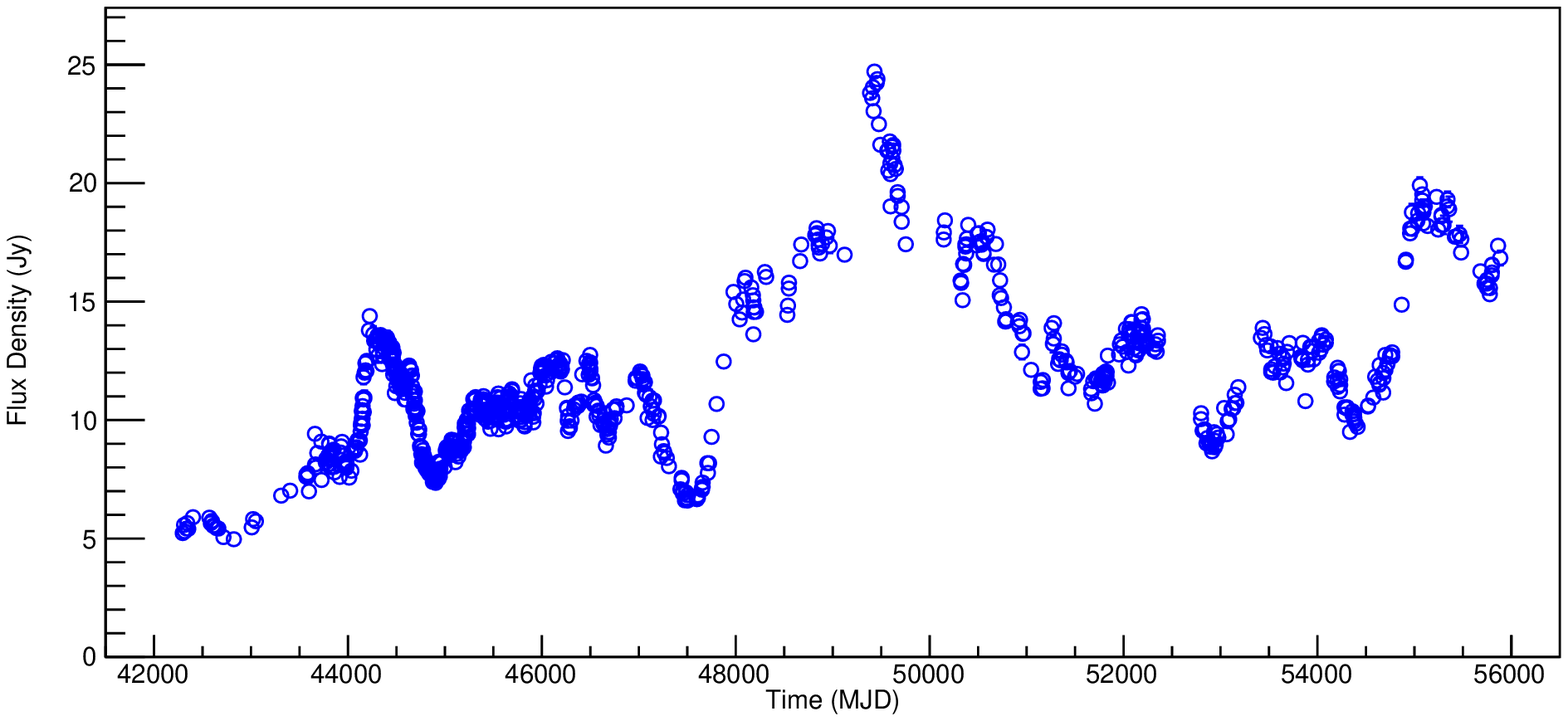}%
\caption{Light curve of PKS~1921-293 radio source in $8.0~\mathrm{GH\lowercase{z}}$, saw the raw dataset.}
\label{fig:rawov236080}
\end{figure}

Likewise, the same preprocessing step give the result shown in the Fig.~\ref{fig:rawov236080} for PKS~1921-293 (OV~236) radio source. In this way, was possible visualize the segments of light curves that had most discontinuities. As before, Figs.~\ref{fig:rawov236048} and \ref{fig:rawov236145} for frequencies $4.8~\mathrm{GH\lowercase{z}}$ and $14.5~\mathrm{GH\lowercase{z}}$ are Appendix~\ref{sec:appPreprocess}.

Notice the time intervals for each radio source shown in Table~\ref{tab:databases} and in the Figs.~\ref{fig:rawdatasetsbllac} in Appendix~\ref{sec:appRawData}, differ from those shown in Fig.~\ref{fig:rawbllac080} and Figs.~\ref{fig:rawbllac048} and~\ref{fig:rawbllac145} as well as from the curves in Fig.~\ref{fig:rawdatasetsov236} in Appendix~\ref{sec:appRawData}, differ from those show in Fig.~\ref{fig:rawov236080} and Figs.~\ref{fig:rawov236048} and~\ref{fig:rawov236145}. Such discrepancy is due to head and tail elimination of the dataset effected during the preprocessing procedure.

\subsection{Regularizing time series}
\label{sec:regseries}

\begin{figure}[!h]
 \centering
 \includegraphics[width=.34\linewidth]{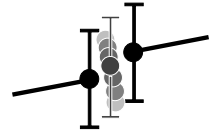}%
 \caption{Schematic representation of how algorithm using XGBoost finds a missing point to fill and complete the time serie irregularly spaced. The error bars represent as the error values can limit the search space.}
 \label{fig:mymethod}
\end{figure}

The UMRAO datasets containing the individual frequencies have several time gaps configuring an irregularly spaced time series. In this method step, XGBoost was used to fill the intervals applying machine learning regression techniques, rather than conventional techniques or methods of usual statistic adjusting. The strategy employed with XGBoost is shown in Fi\-gu\-re~\ref{fig:mymethod}.

In the Fig.~\ref{fig:mymethod}, the black point represents the best point found, i.e., flux density value, to fill the series at that missing point. Gradually darker gray points represent the successive efforts does by new weak learners added to the model to correct the residual errors of all previous trees to choose the point to be tested in the scenario. In this schematic representation, the darker the point, the greater its assigning weights to minimize the loss function.
 
The term regression here refers to the logistic regression or soft-max for the classification task. XGBoost uses a set of decision trees as described the working principle detailed in \S~\ref{sec:xgboost}.

In the interval defined by each error values, $\epsilon_{o}$, were prepared features using moving averages, weighted averages, so that recursive, that is, mean of average, etc., providing XGBoost training phase which several intervals of time below, $\varepsilon_{o-n}$, and above, $\varepsilon_{o+n}$, the point considered. Since that
\begin{equation}
\varepsilon_{o-n}, \ldots, \varepsilon_{o-2}, \varepsilon_{o-1}, \varepsilon_{o}, \varepsilon_{o+1}, \varepsilon_{o+2}, \ldots, \varepsilon_{o+n}\,,
\label{eq:errorts}
\end{equation}
the algorithm search for the best value that can be put at missing point $\varepsilon_{o}$.

The method of classification was to select known points, hide this from algorithm as the training set and the remaining samples in the dataset as the test set (subject to artificial balancing by under-sampling the knowing flux density values). The points chooses by the algorithm fully matches with the previously hided points for the same position in time serie. By this process, applied for all UMRAO radio sources datasets in this study, the irregularly spaced time series, become to regularly spaced time series.

\begin{figure}[!h]\centering
\includegraphics[width=1.0\linewidth]{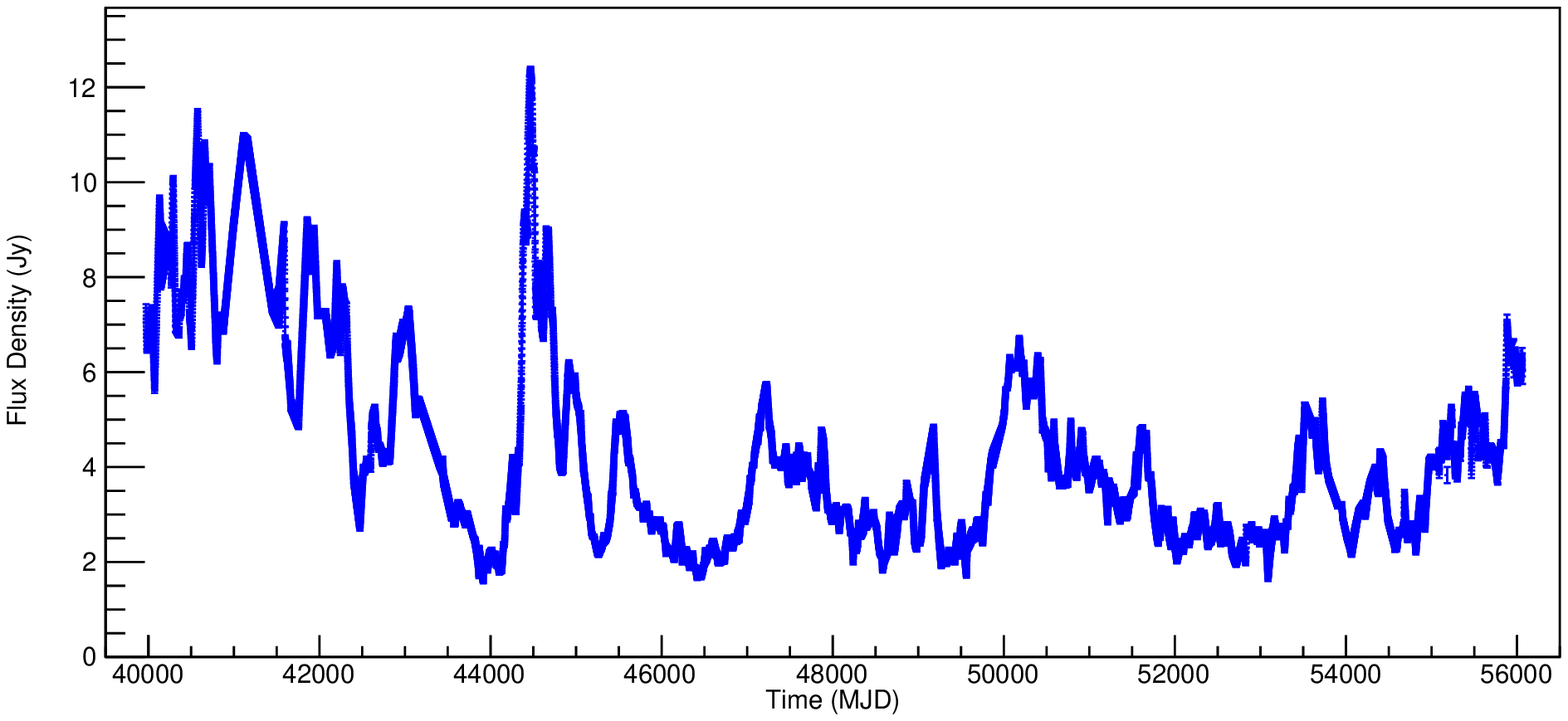}%
\caption{Light curve of PKS~2200+420 (BL~Lac) radio source, in $8.0~\mathrm{GH\lowercase{z}}$, shaw the space-time series regularized.}
\label{fig:regbllac080}
\end{figure}

The Fig.~\ref{fig:regbllac080} present the time series regularized for the for PKS~2200+420 (BL~Lac) radio source in $8.0~\mathrm{GH\lowercase{z}}$ frequency, by the process described. The Appendix~\ref{sec:appRegular} contains Figs.~\ref{fig:regbllac048} and~\ref{fig:regbllac145} with the frequencies $4.8~\mathrm{GH\lowercase{z}}$ and $14.5~\mathrm{GH\lowercase{z}}$.

\begin{figure}[!h]\centering
\includegraphics[width=1.0\linewidth]{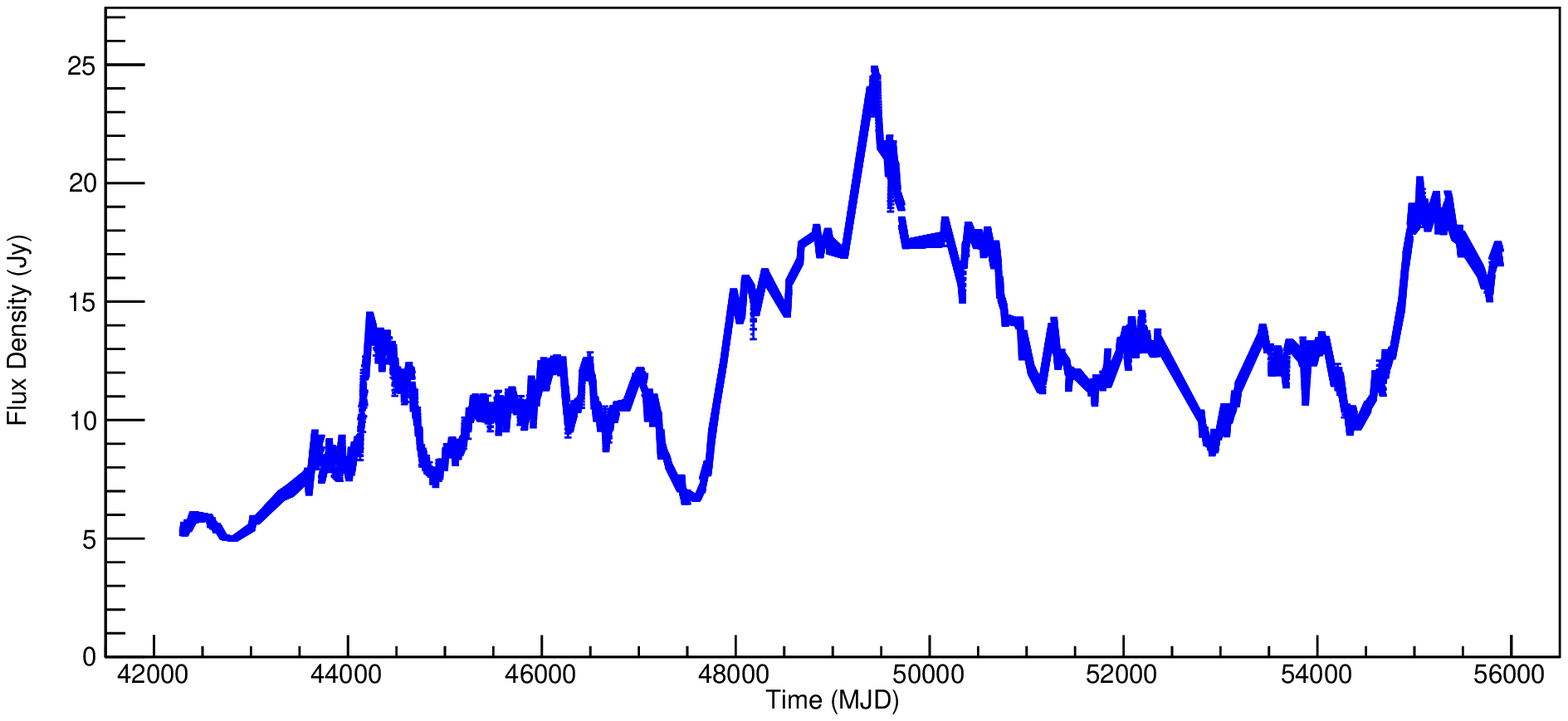}%
\caption{Light curve of PKS~1921-293 (OV~236) radio source in $8.0~\mathrm{GH\lowercase{z}}$, saw the space-time series regularized.}
\label{fig:regov236080}
\end{figure}

Likewise, Fig.~\ref{fig:regov236080} shows the regularized time series for the PKS~1921-293 (OV~236) radio source. The frequencies $4.8~\mathrm{GH\lowercase{z}}$ and $14.5~\mathrm{GH\lowercase{z}}$ frequencies are shown in Figs.~\ref{fig:regov236048} and~\ref{fig:regov236145} in Appendix~\ref{sec:appRegular}.

The accuracy of the regularization of the time series procedure was tested in three steps as follows.
In first step, aproximately one quarter of points, constituted of flux density versus time from the original raw dataset of each frequency, are randomly choose and put separately in differents datasets for the next step.

The file that contain one quarter of the points randomly extracted from the raw dataset are reservated to future comparation. The three quarters of raw data are put in another file to be processed by the algorithm of regularization of time series. The file with three quarters of raw data are processed by the algorithm of regularization of time series, generating an another file with the regularized time serie.

The file with one quarter of the randomly selected points in the first step and the file with the regularized time series generated in the second step are compared. The agreement between the separated and new ones points produced by the regularization algorithm is comparated through the Kolmogorov-Smirnov Test (K-S Test) \citep{Marsaglia2003, Bakoyannis2019, Sadhanala2019, delBarrio2019}.

After dataset regularization by the strategy implemented using XGBoost, any well-established statistical autoregressive model would be conveniently apply over the time series. However, we wanted to experiment and extend the use of XGBoost as much as possible and investigate its capabilities and possibilities of using machine learning also as a tool for time series regularization.

\subsection{Finding outbursts in light curves}
\label{sec:findout}

In this step method, XGBoost was used to classify light curve segments as probably representative of an outburst. It's about, therefore, a binary classification.
First of all, a data segment of light curve containing a knowing outburst was used during the training session. 
Second, in the test session, the light curve of each frequency dataset was modified by the Synthetic Minority Over-sampling Technique, SMOTE~\citep{Chawla2002, Bethapudi2018, HosenieLyon2020}, producing a newest artificial dataset, basically by noise introduction.

By generating a known amount of the simulated noise signal and hiding it among a dataset as if being background noise, one can test if the analysis works, showing that the physics signal of radio source is indeed detectable among the many unknown physics effects. Furthermore, is possible also take note of how many false positives and negatives are introduced in each filtering step and use that data to (\textit{a}) optimize the XGBoost features and (\textit{b}) evaluate the systematic uncertainly of the analysis.

The XGBoost algorithm detected peaks in light curves at all frequencies during training and testing processes using the series artificially created by SMOTE.
This procedure assigns a high degree of confidence to the algorithm, since it was able to identify the same peaks in the artificial and original light curve.

This method step can be sequencing as follow:

\begin{enumerate}
  \item In training and testing processes, the light curve of each frequency already regularized by the previous step (\S~\ref{sec:regseries}). The peak of greater value was taken, assuming that characterizes an outburst.
  \item Preparation features, including samples covering a range of flux density with several time intervals, in days, before and after the occurrence of greatest value peak in each light curve.
  \item In testing process, the algorithm was applied in outburst detection on synthetic datasets created with SMOTE.
  \item Finally, the algorithm was applied on real data sets, looking for segments of the light curve of each frequency as being or not an outburst; doing a binary classification.
\end{enumerate}

\begin{figure}[!h]\centering
\includegraphics[width=1.0\linewidth]{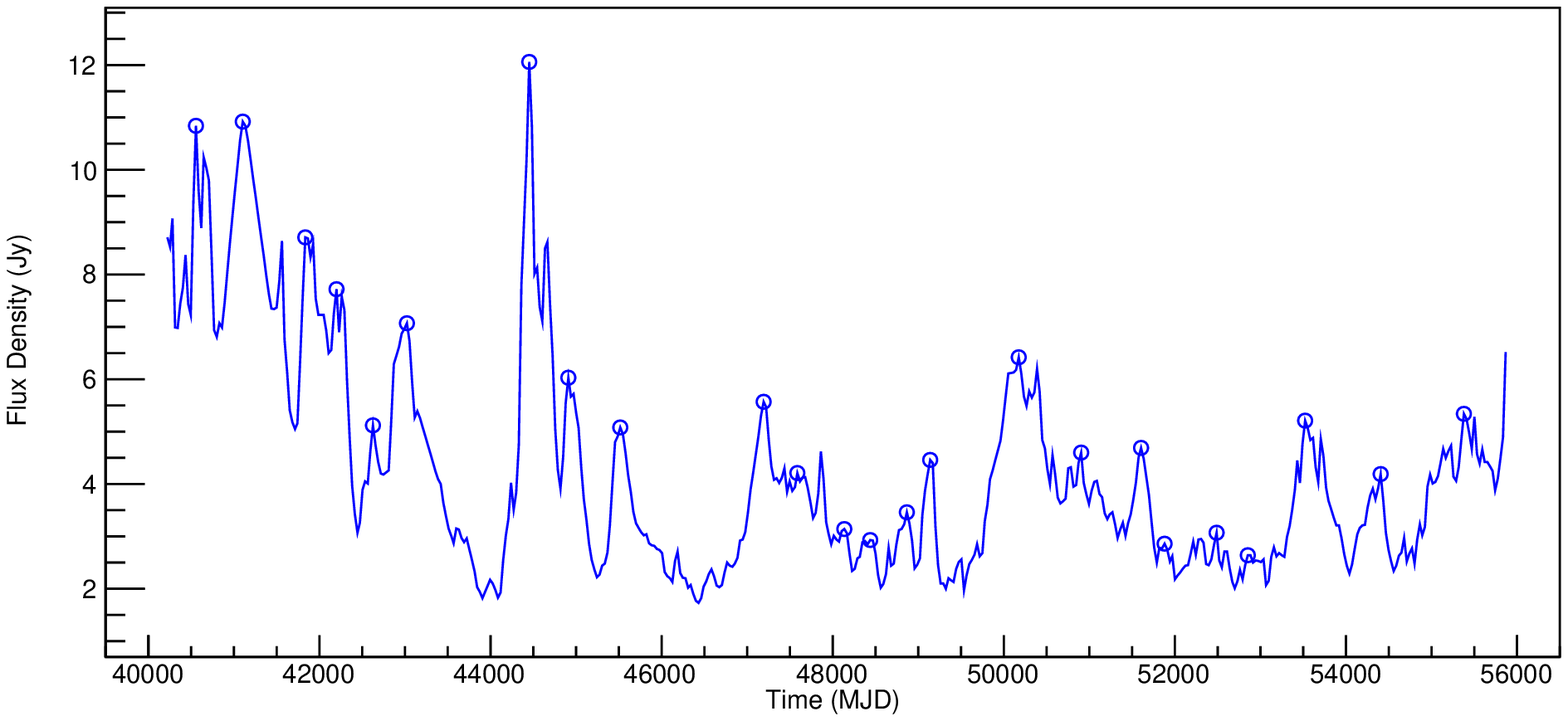}%
\caption{Light curve of PKS~2200+420 (BL~Lac) radio source, in $8.0~\mathrm{GH\lowercase{z}}$, shaw the detected outbursts to found periodicities.}
\label{fig:peakbllac080}
\end{figure}

At the end of this step, the outburst candidates detected by the algorithm for PKS~2200+420 (BL~Lac) radio source, in $8.0~\mathrm{GH\lowercase{z}}$, were plotted in graph, as shown in the Fig.~\ref{fig:peakbllac080}. All other frequencies for this object are in Figs.~\ref{fig:peakbllac048} and~\ref{fig:peakbllac145}, Appendix~\ref{sec:appOutBurst}.

\begin{figure}[!h]\centering
\includegraphics[width=1.0\linewidth]{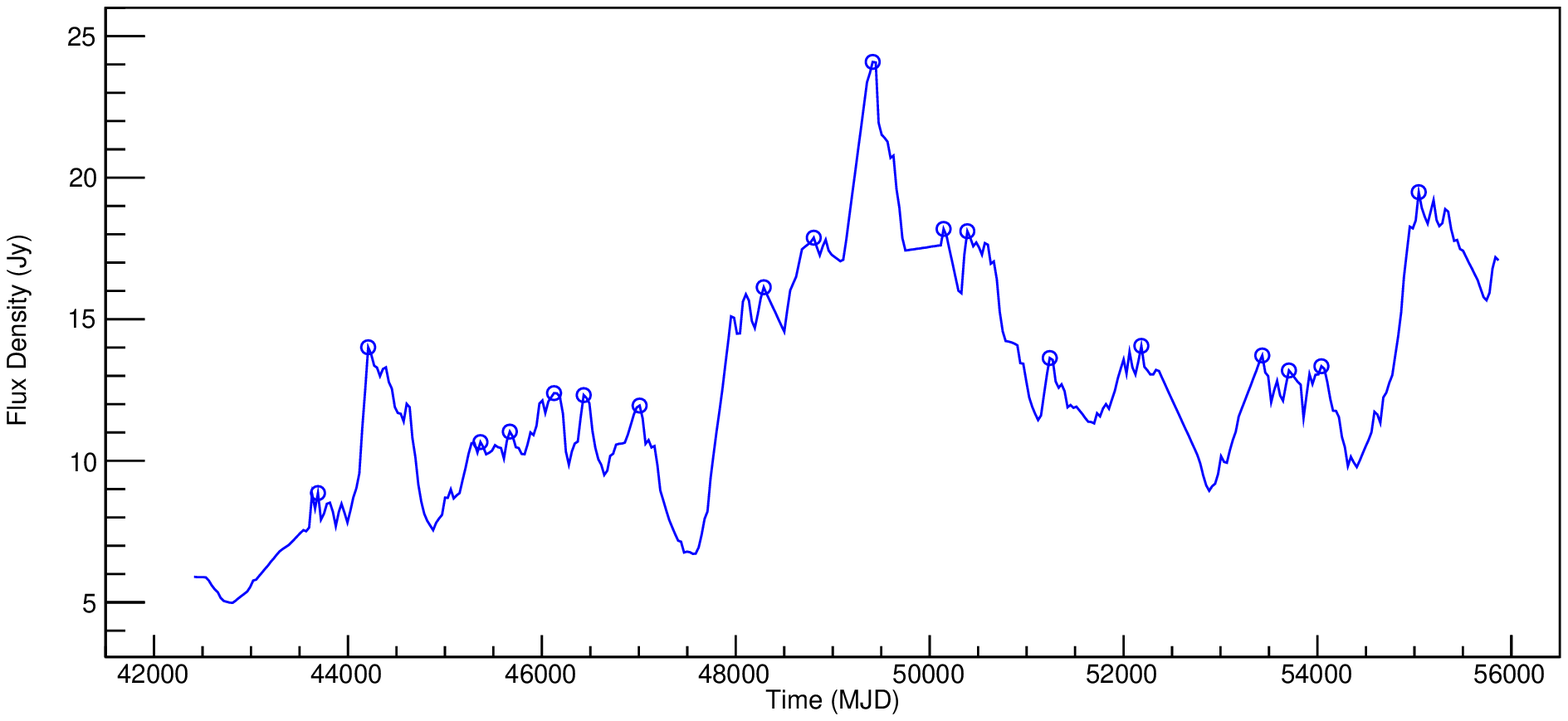}%
\caption{Light curve of PKS~1921-293 (OV~236) radio source, in $8.0~\mathrm{GH\lowercase{z}}$, saw the the detected outbursts to found periodicities.}
\label{fig:peakov236080}
\end{figure}

Thus also for PKS~1921-293 (OV~236) radio source in the Fig.~\ref{fig:peakov236080} and Figs.~\ref{fig:peakov236048} and \ref{fig:peakov236145}, Appendix~\ref{sec:appOutBurst}.

\subsection{Periodicity}
\label{sec:periodicity}

The calculation to determine the periodicity was take the combination of the differences of time among all the outburst candidates identified in the classification by XGBoost algorithm in the previous method step (\S~\ref{sec:findout}). The goal was to determine all possible combinations between the occurrences of outbursts by collecting the corresponding time intervals.

Each outburst candidate corresponds to an ordered pair of flux density with an occurrence date.
The difference between all possible occurrence date combinations of all outburst candidates provides a set of time intervals within which may contain the periodicity of the phenomenon within the appropriate boundary conditions. This boundary conditions were proposed by \citet{RasheedAlshalalfaAlhajj2011}.

At this point, it is necessary to define what was considered `periodicity' within the scope of this research.

\citet{RasheedAlshalalfaAlhajj2011} distinguish between seven different definitions of periodicity. The one that interests in our context is the Periodicity with Time Tolerance. This postulates that, given a time series $T$ which is not necessarily noise-free, a pattern $X$ is periodic in an interval $[\mbox{startPos}, \mbox{endPos}]$ of $T$ with period $p$ and time tolerance $tt\geq 0$, if $X$ is found at positions
\begin{equation}\footnotesize
\mbox{startPos} + p \pm tt, \mbox{startPos} + 2p \pm tt, \ldots , \mbox{endPos} + p \pm tt\,.
\label{eq:pattern}
\end{equation}

Because it is not always possible to achieve perfect periodicity and hence we need to specify the confidence in the reported result. \citet{RasheedAlshalalfaAlhajj2011} define the periodicity confidence, as follow.

The confidence of a periodic pattern $X$ occurring in time series $T$ is the ratio of its actual periodicity to its expected perfect periodicity.

Formally, the confidence of pattern $X$ with periodicity $p$ starting at position $\mbox{startPos}$ is defined as:
\begin{equation}\footnotesize
\mathtt{conf}\left( p, \mbox{startPos},X \right) = \frac{ P_{\texttt{Actual}}\left( p, \mbox{startPos}, X \right) }{ P_{\mbox{Perfect}}\left( p, \mbox{startPos}, X \right) }
\label{eq:confidence}
\end{equation}
where the perfect periodicity is,
\begin{equation}\footnotesize
P_{\mbox{Perfect}} \left( p, \mbox{startPos}, X \right) = \left[ \frac{ \lvert T \rvert - \mbox{starPos} + 1}{p} \right]
\label{eq:perfectperiod}
\end{equation}
and the actual periodicity $P_{\texttt{Actual}}$ is calculated by counting the number of occurrences of $X$ in $T$, starting at $\mbox{startPos}$ and repeatedly jumping by $p$ positions.

Thus, for example, in $T = abbcaabcdbaccdbabbca$, the pattern $ab$ is periodic with $\mbox{startPos} = 0$, $p = 5$, and $\mathtt{conf}(5, 0, ab) = 3/4$. Note that the confidence is $4/4 = 1$ when perfect periodicity is achieved.

The correspondence between the time series, $T$, and the chain of binary digits, in which the '1's mark the position of the periodic pattern $X$ occurrence in the series, helps to clarify the definition.

\begin{verbatim}
abbcaabcdbaccdbabbca
10000100000000010000
\end{verbatim}

Applying the confidence definition (Eq.~\ref{eq:confidence}) in Periodicity with Time Tolerance like $T = abce~dabc~cabc~aabc~babc~c$, the frequency is $\mathtt{freq}(ab, 4, 0, 18, tt = 1) = 5$ and the confidence is $\mathtt{conf}(ab, 4, 0, 18, tt = 1) = 5/5 = 1$ \citep{RasheedAlshalalfaAlhajj2011}.

The concepts as defined here were used to compute and validate the periodicities found in the datasets, $T$, of the radio sources examined.

The outbursts represent the periodic pattern, $X$. The time tolerance, $tt$, assumed was that of the arithmetic mean difference in days between the arrival times, to the observer, of the main outbursts at each frequency (Eq.~\ref{eq:timetolerance}).

\begin{equation}
tt = \frac{\Delta t_{|f1 - f2|} + \Delta t_{|f1 - f3|} + \Delta t_{|f2 - f3|}}{3}
\label{eq:timetolerance}
\end{equation}

This way of stipulating the time tolerance $tt$ is based on the \textit{ansatz} that any real comparison or correlation between two or more radio sources frequencies examined, must take into account the temporal separation between the incoming of the characteristic peaks of the outbursts to the observer point of view.

This way of stipulating $tt$ is as expressed in Eq.~(\ref{eq:pattern}) to a temporal interval.

\begin{figure}[!ht]
  \centering
  \includegraphics[width=1.0\linewidth]{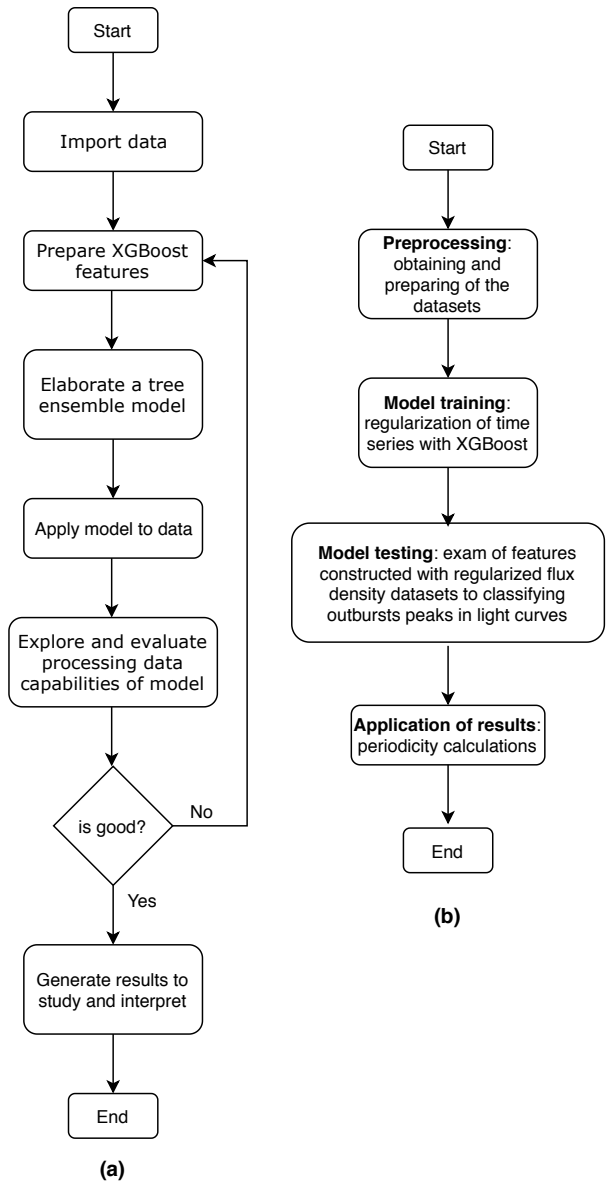}%
  \caption{In \textbf{(a)} the typical algorithm of machine learning to which, whatever the purpose, any task ends up being reduced. In \textbf{(b)} Sequences of the general method employed. Each box in diagram \textbf{(b)} may have one or more steps as described in \textbf{(a)}.}
  \label{fig:method}
\end{figure}

The Fig.~\ref{fig:method} shows a synthesis diagram of the steps of the applied method, emphasizing how the aspects inherent to the use of the XGBoost and those referring to the data and to the phenomenon studied were contemplated in the design of the method.

As in the case of the methodological option assumed in the regularization of time series in \S~\ref{sec:regseries},
here too we have chosen to explore the limits of possibilities without making use of traditional methods of calculating periodicity. For this reason, we have adopted the postulates proposed by \citet{RasheedAlshalalfaAlhajj2011} to compute and validate the periodicities.

\section{Results and Discussion}
\label{sec:results}

The initial methodological step, presented in \S~\ref{sec:prepro}, eliminated flux density data temporally very far from each other only in the head and tail of datasets. This process was necessary to reduce computational time of next steps, otherwise, mainly the time series regularization would be harder for the algorithm do it. As result, the time series was shortened by a few days on his head and tail. This did not bring perceptible losses to the accuracy of the method since the time series datasets were very extensive. At the head of the time series, the dates eliminated were in November of one year and the next one was in February the following year. In the tail of the time series, the days of the following year were eliminated, closing them in the final days of the previous year. Eliminating these days from the heads and tails of the time series allowed us to balance the computational time/accuracy ratio.

After refining and tuning the datasets as described above, separating frequencies into distinct files to prepare the XGBoost features was the next processing.

The strategy used in the next step of the method is to make the XGBoost-based algorithm consider past and future events of flux density, to weight and predict which flux density value within the same frequency examined, most suitable to be placed at a point missing between the past and future points of the time series.

The features delivered to XGBoost for processing in this step were prepared as a ``sliding window'', that traverses the points that make up the light curves of each frequency one by one (increment 1) and repeating the process with each increment.

Some known points (originally existing in the time series) were hidden and the algorithm was asked to compute its value without knowing it previously as described previously. The results of $p$ values can vary slightly due to the fact that the selected randomly points change with each execution of the algorithm that calculate the K-S test. Even so, the $p$ values variation does not escape from aproximately 99\% as summarized in Table~\ref{tab:KSvalues}.

\begin{table*}[ht]\centering
    \setlength{\tabnotewidth}{0.8\linewidth}
    \setlength{\tabcolsep}{0.9\tabcolsep} \tablecols{8}
  \caption{Result of the K-S Test applied on two samples. One with the raw empirical data of flux density and the another calculated, for the same point, with the algorithm of regulation based on tree boosting.}
  \label{tab:KSvalues}
  \begin{tabular}{r r r c r r}
   \toprule
   & \multicolumn{2}{c}{PKS~2200+420 (BL~Lac)} && \multicolumn{2}{c}{PKS~1921-293 (OV~236)} \\
   \cmidrule{2-3}\cmidrule{5-6}
   Frequency & statistic & $p$ value && statistic & $p$ value \\
   \midrule
   $4.8~\mathrm{GH\lowercase{z}}$ & $0.03571429$ & $0.99989572$ && $0.03246753$ & $0.99999660$ \\
   $8.0~\mathrm{GH\lowercase{z}}$ & $0.03349282$ & $0.99974697$ && $0.03097345$ & $0.99988705$ \\
 $14.5~\mathrm{GH\lowercase{z}}$ & $0.03361345$ & $0.99913286$ && $0.02734375$ & $0.99997300$ \\
   \bottomrule
  \end{tabular}            
\end{table*}

The empirical data \texttt{versus} values calculated by the time series regularization algorithm have a very high statistical adherence of about $0.99$. Thus, the K-S Test rejects the the null hypothesis that the samples are drawn from the same distribution in the two-sample case. It means that there is no evidence to say that the set of values does not adhere, so it is understood then that the calculated values come from the same probability distribution, since it is a very high correlation in all cases. In another words, the probability of these two sample not coming from same distribution is very low. But, statistically speaking, can not be 100\% sure about that. 

For this reason it is used the method of regularization based on tree boosting instead of conventional techniques.
This process offers more convincing results as to belonging to a natural phenomena than the use of more conventional techniques such as spline, for example, that make too much smooth the curves from experimental data.

What is seen in the curve shown in Fig.~\ref{fig:regbllac080} (\S~\ref{sec:regseries}) and Figs.~\ref{fig:regbllac048} and~\ref{fig:regbllac145} (Appendix~\ref{sec:appRegular}) of radio source PKS~2200+420 (BL~Lac) and Fig.~\ref{fig:regov236080} and Figs.~\ref{fig:regov236048} and~\ref{fig:regov236145} of radio source PKS~1921-293 (OV~236), are actually plots of daily points, merging previously obtained points from UMRAO and points forecasted by the XGBoost algorithm. The case is that: this is not a line in fact. The tessellated aspect results from the fact that some values of the flux density (real or forecasted) are much higher or lower than their predecessors or successors.

Table~\ref{tab:hyperregular} summarize the hyper parameters values with which XGBoost was configured in this step of the method to obtain the results.

\begin{table}[h]
  \centering
  \caption{Hyper parameters of XGBoost models used for method applied for time series regularization discussed here.}
  \label{tab:hyperregular}
  \begin{tabular}{l r}
   \toprule
   Number of estimators & 1000 \\
   Learning rate  & 0.001 \\
   Maximum depth  & 4 \\
   Regularization alpha & 0.01 \\
   Gamma & 0.1 \\
   Sub sample & 0.8 \\
   \bottomrule
  \end{tabular}            
\end{table}

XGBoost Python API\footnote{All information needed to properly install and use XGBoost Python API is available at: \url{https://xgboost.readthedocs.io/en/latest/python/python_intro.html}.} provides a method to assess the incremental performance by the number of trees. It uses arguments usually to train, test and to measure errors on these evaluation sets. This allowed us to adjust the model performance until get the best result in terms computational time/accuracy ratio presented herein. 

In Table~\ref{tab:hyperregular}, as in Table~\ref{tab:hyperoutburst}, the XGBoost hyper parameters are~\citep{Aarshay2016}:
\begin{description}
  \item [Number of estimators] sets the number of trees in the model to be generated.
  \item [Learning rate] affects the computation time performance that decreases incrementally the learning rate while increasing the number of trees.
  \item [Maximum depth] represents the depth of each tree, which is the maximum number of different features used in each tree.
  \item [Regularization alpha] is the linear booster term on weight and controls the complexity of the model which prevents overfitting,
  \item [Gamma] specifies the minimum loss reduction required to make a node splitting in the tree. This occurs only when the resulting split gives a positive reduction in the loss function.
  \item [Sub sample] is the percentage of rows obtained to build each tree. Decreasing it, reducing performance.
\end{description}

Good performance of XGBoost-based algorithms depends on the ability to adjust the hyper parameters of the model.
It took time of processing in the training and testing processes, after preparing the models features, to achieve the results shown in Fig.~\ref{fig:regbllac080} and Figs.~\ref{fig:regbllac048} and~\ref{fig:regbllac145} for PKS~2200+420 (BL~Lac) datasets and Fig.~\ref{fig:regov236080} and Figs.~\ref{fig:regov236048} and~\ref{fig:regov236145} for PKS~1921-293 (OV~236) datasets.

The algorithms XGBoost based emphatically exhibited potential ability for outbursts detection in radio sources light curves.
Even when the datasets were disturbed with artificial noise introduction by SMOTE, the XGBoost algorithm stay the ability of outbursts identification, matching with the previous findings in all frequencies, without noise. The robustness of the method and the solid boosted tree implementation behind the algorithms are validated by the proximity of the scores computed for different datasets of both radio sources.

Fig.~\ref{fig:peakbllac080} and Figs.~\ref{fig:peakbllac048} and~\ref{fig:peakbllac145} of radio source PKS~2200+420 (BL~Lac) and
Fig.~\ref{fig:peakov236080} and Figs.~\ref{fig:peakov236048} and \ref{fig:peakov236145} of radio source PKS~1921-293 (OV~236), show the peaks identified by the algorithm XGBoost-based, according to the methodological strategy described in \S~\ref{sec:periodicity}. 

Table~\ref{tab:hyperoutburst} summarize the hyper parameters values which XGBoost was configurated in this step of the method. The hyper parameters and were the same in both, SMOTE simulated and non-simulated with acquired UMRAO data. This strategy was inspired by \citet{Bethapudi2018}.

\begin{table}[!ht]
  \centering
  \caption{Hyper parameters of XGBoost models used for method applied for outburst detection. We have used the same set of hyper parameters for both the non-SMOTE and SMOTE datasets.}
  \label{tab:hyperoutburst}
  \begin{tabular}{l r}
   \toprule
   Number of estimators & 200 \\
   Learning rate  & 0.001 \\
   Maximum depth  & 10 \\
   Regularization alpha & 0.0001 \\
   Gamma & 0.1 \\
   Sub sample & 0.6 \\
   \bottomrule
  \end{tabular}            
\end{table}

As in the previous methodological step, at this stage also the skills to adjust the hyper parameters was essential to obtain the results discussed here.

The use of previously adjusted datasets contributed to gain precision in the accuracity detection of the outbursts, since it enlarged the sample space and the temporal resolution.

XGBoost, as well as other implementations of tree optimization algorithms, is a good choice for both classification and prediction.
But decision tree ensembles models is not directly aplicable for variability or periodicity studies. Some kind of smarth strategy was requered to be able extract periodicity from light curves using tree boosting.

In fact, the XGBoost contribution to periodicity calculation ended with the identifies of outburst candidates from light curves. Thereafter, the method is reduced to calculate differences, subtracting all pairs of peaks found from each other and verifying if the values found fall within a time tolerance limit.

The algorithm based on XGBoost was submitted to two tests. In the first, artificial datasets SMOTE were used to verify if the algorithm would find the candidate peaks to outburst despite the introduced a random combination of artificial noises in data sets by the SMOTE technique. In the second, the light curves were inverted in such a way that the first point of the curve became the last and vice versa. After re-training the algorithm, all points were identified in both cases.

Finally. after the training and test procedures, the algorithm was applied to the UMRAO datasets of both object, obtaining good results.

The obtained results were compared with others, from previous works obtained using the same datasets, but different statistical methods. In addition to the difference in method and size of the time series, which were smaller than the time series used in this work, since they were from years ago and, therefore, the datasets used here were not available, a notice characteristic the works consulted, is that employed conventional ways to treat irregular time series.

The results for PKS~2200+420, shown in Table~\ref{tab:perbllac}, are compared with the results found in the several works, collected in Table~\ref{tab:comparebllac} for methods
Discrete Fourier Transform, discrete AutoCorrelation Function (DFT/ACF) \citet{Villata2004}, Simultaneous Threshold Interaction Modeling Algorithm (STIMA) \citet{Ciaramella2004},
Power Spectral Analysis Method (PSA) \citet{Yuan2011}, Date-Compensated Discrete Fourier Transform (DCDFT) \citet{FanLiu2007} and 
Continuous Wavelet Transform, Cross-Wavelet Transform (WT) \citet{KellyHughes2003}.

\begin{table}[!ht]
  \centering
  \caption{Periodicity obtained for the PKS~2200+420 (BL~Lac) radio source after computational procedure to classify flux density segments as potential outburst. Values with a precision of $88.95\%$.}
  \label{tab:perbllac}
  \begin{tabular}{r c c}
   \toprule
   Frequency & Time interval & Periodicity (year) \\
   \midrule
     $4.8~\mathrm{GH\lowercase{z}}$ & $1978$--$2011$ & $1.7$, $3.4$, $5.7$ \\
     $8.0~\mathrm{GH\lowercase{z}}$ & $1969$--$2011$ & $1.7$, $3.8$, $5.2$ \\
   $14.5~\mathrm{GH\lowercase{z}}$ & $1975$--$2011$ & $1.7$, $2.9$, $4.7$ \\
   \bottomrule
  \end{tabular}            
\end{table}
\begin{table*}[!ht]
  \centering
  \caption{Comparison of the periodicity obtained for PKS~2200+420 with periodicities estimated by different methods.}
  \label{tab:comparebllac}
  \begin{tabular}{c l l l l}
   \toprule
   Time interval & Method & $4.8~\mathrm{GH\lowercase{z}}$ & $8.0~\mathrm{GH\lowercase{z}}$ & $14.5~\mathrm{GH\lowercase{z}}$ \\
   \midrule
   $1968$--$2003$ & DFT/ACF & $1.4~\mathrm{yr}$                & $3.7~\mathrm{yr}$                & $7.5; 1.6; 0.7~\mathrm{yr}$  \\
   $1977$--$2003$ & STIMA     & $7.8~\mathrm{yr}$                & $6.3~\mathrm{yr}$                & $7.8~\mathrm{yr}$  \\
   $1968$--$1999$ & PSA          & $5.4; 9.6; 2.1~\mathrm{yr}$ & $4.9; 9.6; 2.8~\mathrm{yr}$ & $2.4; 4.3. 14.1~\mathrm{yr}$ \\
   $1977$--$2005$ & DCDFT     & $3.9; 7.8~\mathrm{yr}$        & $3.8; 6.8~\mathrm{yr}$         & $3.9; 7.8~\mathrm{yr}$ \\
   $1984$--$2003$ & WT           & $1.4~\mathrm{yr}$                & $3.7~\mathrm{yr}$                & $3.5; 1.6; 0.7~\mathrm{yr}$ \\
   \bottomrule
  \end{tabular}
\end{table*}

The results for PKS~1921-293, shown in Table~\ref{tab:perquasar}, are compared with the results found in the~\citet{Gastaldi2016} work, collected in Table~\ref{tab:comparequasar}, unlike the various works collected to compare with the result to PKS~2200+420 (BL~Lac). It is rasonable why~\citet{Gastaldi2016} made a full review in his phd thesis about others methods of find out periodicity to comapar with his own method to calculate periodicity of PKS~1921-293.

When comparing the results of Table~\ref{tab:perbllac} with the Table~\ref{tab:comparebllac} and of Table~\ref{tab:perquasar} with Table~\ref{tab:comparequasar}, it is noted that they are similar to each other. 
It is recommended to keep in mind that the time series intervals are different and smaller than those used in this paper. In spite of this, and of the methods used in the manner in which the data are processed, it is seen that the periods approach, in particular those of the frequency $14.5~\mathrm{GH\lowercase{z}}$ of the PKS~2200+420 source radio.

These results can only be considered compatible if a time tolerance limit is assumed, estimated through the arrival delay of the maximum peaks of the several frequencies at the observer point of view. The value of this delay for PKS~1921-293 radio source, is, on average, approximately $42$ days, and circa $21$ days for PKS~2200+420.

\begin{table}[!ht]
  \centering
  \caption{Periodicity obtained for the PKS~1921-293 (OV~236) radio source after computational procedure to classify flux density segments as potential outburst. Values with a precision of $89.83\%$.}
  \label{tab:perquasar}
  \begin{tabular}{r c c}
   \toprule
  Frequency & Time interval & Periodicity (year) \\
   \midrule
     $4.8~\mathrm{GH\lowercase{z}}$ & $1980$--$2011$ & $1.2$, $3.6$, $5.0$ \\
     $8.0~\mathrm{GH\lowercase{z}}$ & $1975$--$2011$ & $1.3$, $2.8$, $5.2$ \\
   $14.5~\mathrm{GH\lowercase{z}}$ & $1976$--$2011$ & $1.6$, $3.2$, $6.3$ \\
   \bottomrule
  \end{tabular}
\end{table}
\begin{table*}[!ht]
  \centering
  \caption{Periodicity obtained for the PKS~1921-293 (OV~236) with periodicities estimated by Lomb periodogram and Wavelet methods.}
  \label{tab:comparequasar}
  \begin{tabular}{r c l l}
   \toprule
   Frequency & Time interval & Method & Periodicity (year) \\
   \midrule
   $4.8~\mathrm{GH\lowercase{z}} $ & $1980$--$2006$ & Lomb    & $1.8$, $3.3$, $9.5$ \\
    ~ &  ~ & Wavelet & $1.2$--$1.9$, $2.7$--$2.8$, $5.2$--$5.3$ \\
   $8.0~\mathrm{GH\lowercase{z}} $ & $1981$--$2006$ & Lomb    & $1.3$, $2.8$, $3.0$, $5.0$, $8.5$u \\
   ~ & ~ & Wavelet & $1.2$--$1.4$, $2.3$--$2.6$, $3.2$, $4.3$--$5.1$ \\
   $14.5~\mathrm{GH\lowercase{z}} $  & $1982$--$2006$ & Lomb     & $1.3$, $2.5$, $4.3$, $6.5$ \\
   ~ &  ~ & Wavelet & $1.3$--$2.3$, $3.6$, $5.0$--$5.5$ \\
   \bottomrule
  \end{tabular}
\end{table*}

\section{Summary and conclusions}
\label{sec:summconc}

In order to implement, test and improve a method that incorporates the tree boosting-based machine learning algorithm (XGBoost) for analysis and study of characteristics of radio sources, to figure out the potential capabilities that specific tool have for astrophysics purposes, two typical radio sources datasets of radio sources are explored in the form of time series.
The objects chosen were PKS~1921-293 (OV~236) and PKS~2200+420 (BL-Lac), because they were the most studied in the radio range in many research works and therefore, several attempts to discover the periodicity were performed by different methods. For this, the datasets from University of Michigan Radio Astronomy Observatory (UMRAO), which operates at frequencies of $4.8~\mathrm{GH\lowercase{z}}$, $8.0~\mathrm{GH\lowercase{z}}$ and $14.5~\mathrm{GH\lowercase{z}}$ was chosen.

A method boosting-based algorithms was experimented. The method consists of using XGBoost in two different steps. In the first this machine learning library was exploited in its potential to act as a regression tool and thus to regularize unspaced temporal series, making them regularly space-time. In the second, the potential of XGBoost as a classification tool was emphasized to select regions in the light curves that mark outbursts. In both cases the researcher expertise is an indissociable component of the greater or lesser success of the methodological process.

The XGBoost shows precise probabilistic results, as long as the researcher has a good understanding of the problem and clearly specifies the characteristics of the phenomenon to be studied through well-defined boundary conditions and a validity and tolerance interval of the well-established specified values in the features.

The success or failure of using XGBoost-based algorithms depends on the researcher's skills to adjust the hyperparameters of the model.
It should be noted that XGBoost does not can be used itself to periodicity detection or calculation such as some statistical methods or other Fourier derivative methods. The method used the strategy of the classify outbursts in the light curve, task viable for XGBoost, and later calculated the temporal difference between the candidates to outburst identified by XGBoost classification, of course, from a confidence interval that established the precision and thus, even with discrepancies falling within the interval, find periodic values.

The results found were quite close to those found by other more orthodox methods. Despite this they have the advantage of low computational time, potential to be applied in big data.

In this first approximation of XGBoost with astrophysics through the study of the radio sources, the enormous potential of applications of this algorithm and the machine learning in general for this field of study was perceived. The presented results, by themselves, justify investigating other potential uses for this tool.

Future perspectives involve the extension of the study for other energy ranges, such as X-rays and gamma rays and the exploration of the use of methods based on tree boosting and other machine learning techniques that allow the application in multifrequency analysis .

It is also expected to associate the tree boosting with XGBoost with the Monte Carlo technique to evaluate how well available models are able to describe energy regimes, variability, and other aspects of radio sources.

\vskip 10mm
\section*{Acknowledgments}

We would like to thank financial support from Mackenzie Presbyterian University.
This research has made use of data from the University of Michigan Radio Astronomy Observatory which has been supported by the University of Michigan and by a series of grants from the National Science Foundation, most recently AST-0607523.
Special thanks to Margo F. Aller and Hugh D. Aller for the datasets that made possible this work.

\begin{appendices}

This appendices contains supplementary material that is important part of the research itself, and therefore may be useful in providing a more comprehensive understanding of the work, however is too cumbersome to include in the body of the paper.

\section{Raw dataset of objects PKS~2200+420 and PKS~1921-293}\label{sec:appRawData}

\begin{figure}[!h]
 \centering
 \includegraphics[width=1.0\linewidth]{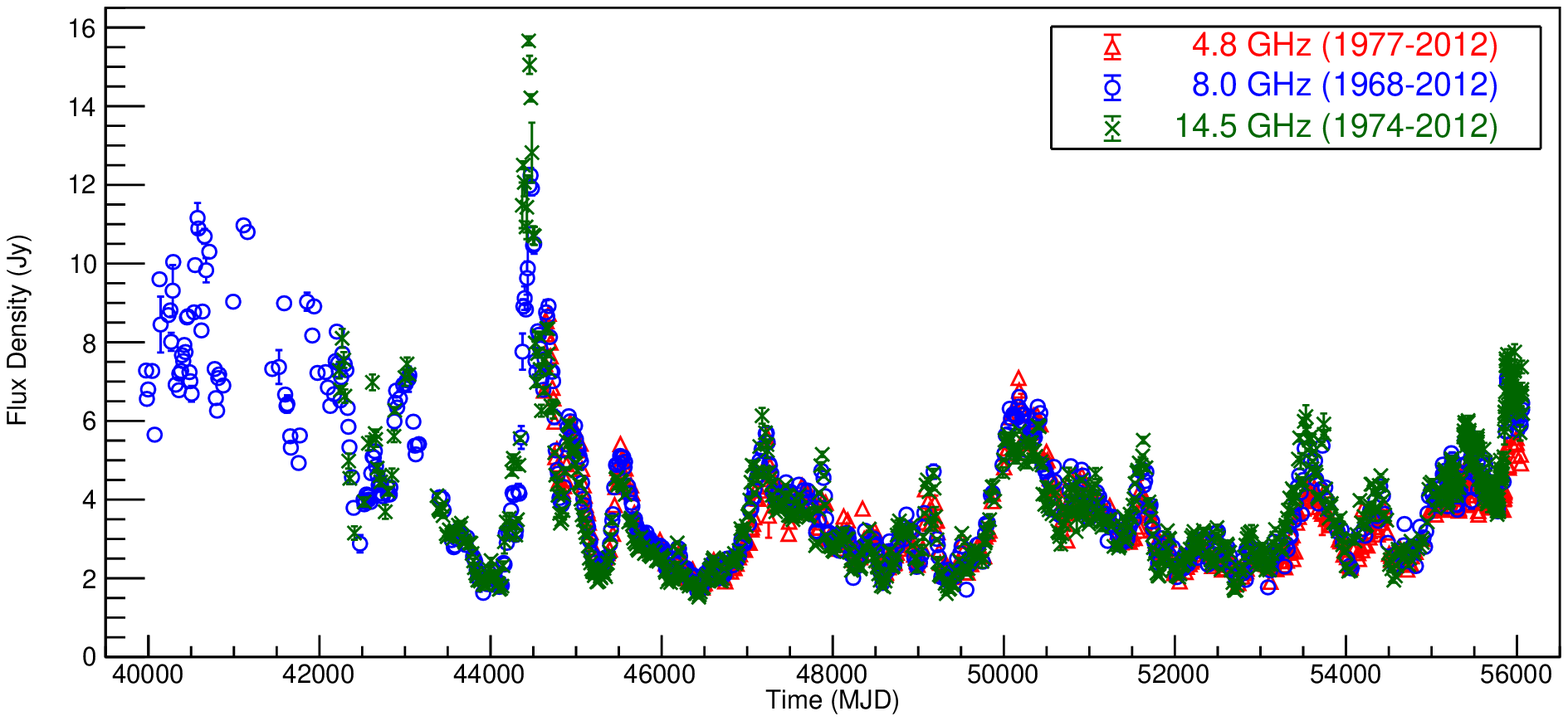}%
 \caption{Raw dataset of object PKS~2200+420 (BL~Lac), made available by UMRAO before adjustments. Note that several point segments in the time series are missing at all disponible frequencies.}
 \label{fig:rawdatasetsbllac}
\end{figure}

\begin{figure}[!h]
 \centering
 \includegraphics[width=1.0\linewidth]{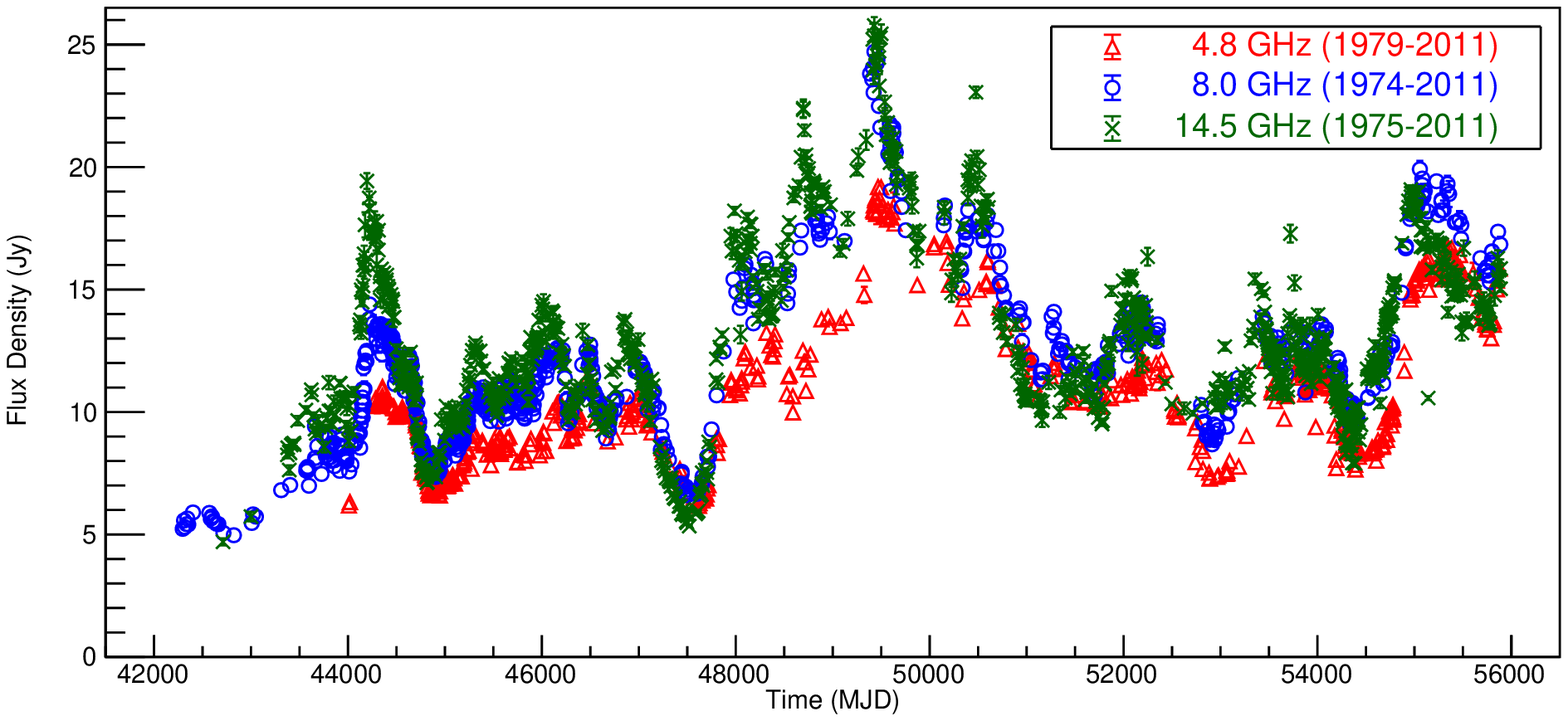}%
 \caption{Raw dataset of object PKS~1921-293 (OV~236), made available by UMRAO before adjustments. Note that several point segments in the time series are missing at all disponible frequencies.}
 \label{fig:rawdatasetsov236}
\end{figure}

The Fig.~\ref{fig:rawdatasetsbllac} and~\ref{fig:rawdatasetsov236} show the datasets of the two objects, PKS~2200+420 (BL~Lac) and PKS~1921-293 (OV~236) respectively, as made available by UMRAO  before adjustments of preprocessing. For clarity the original colors used by UMRAO were maintained in this appendix, as in whole paper, according to the frequencies: red for $4.8~\mathrm{GH\lowercase{z}}$, blue for $8.0~\mathrm{GH\lowercase{z}}$ and green for $14.5~\mathrm{GH\lowercase{z}}$.

\section{Raw dataset of objects PKS~2200+420 and PKS~1921-293 after preprocessing}\label{sec:appPreprocess}

\begin{figure}[!h]\centering
\includegraphics[width=1.0\linewidth]{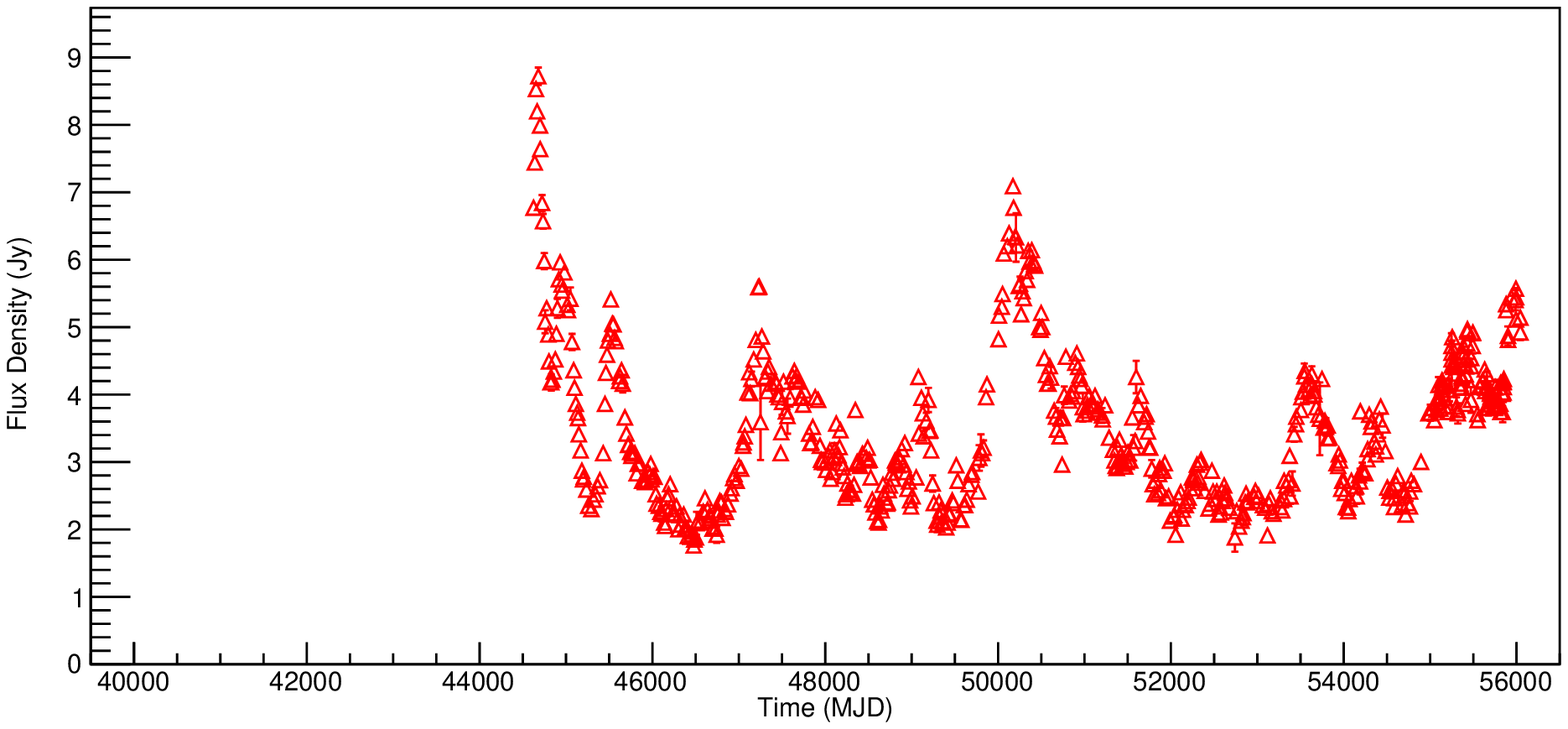}%
\caption{\footnotesize Light curve of PKS~2200+420 radio source, in $4.8~\mathrm{GH\lowercase{z}}$, shaw the raw dataset.}
\label{fig:rawbllac048}
\end{figure}

\begin{figure}[!h]\centering
\includegraphics[width=1.0\linewidth]{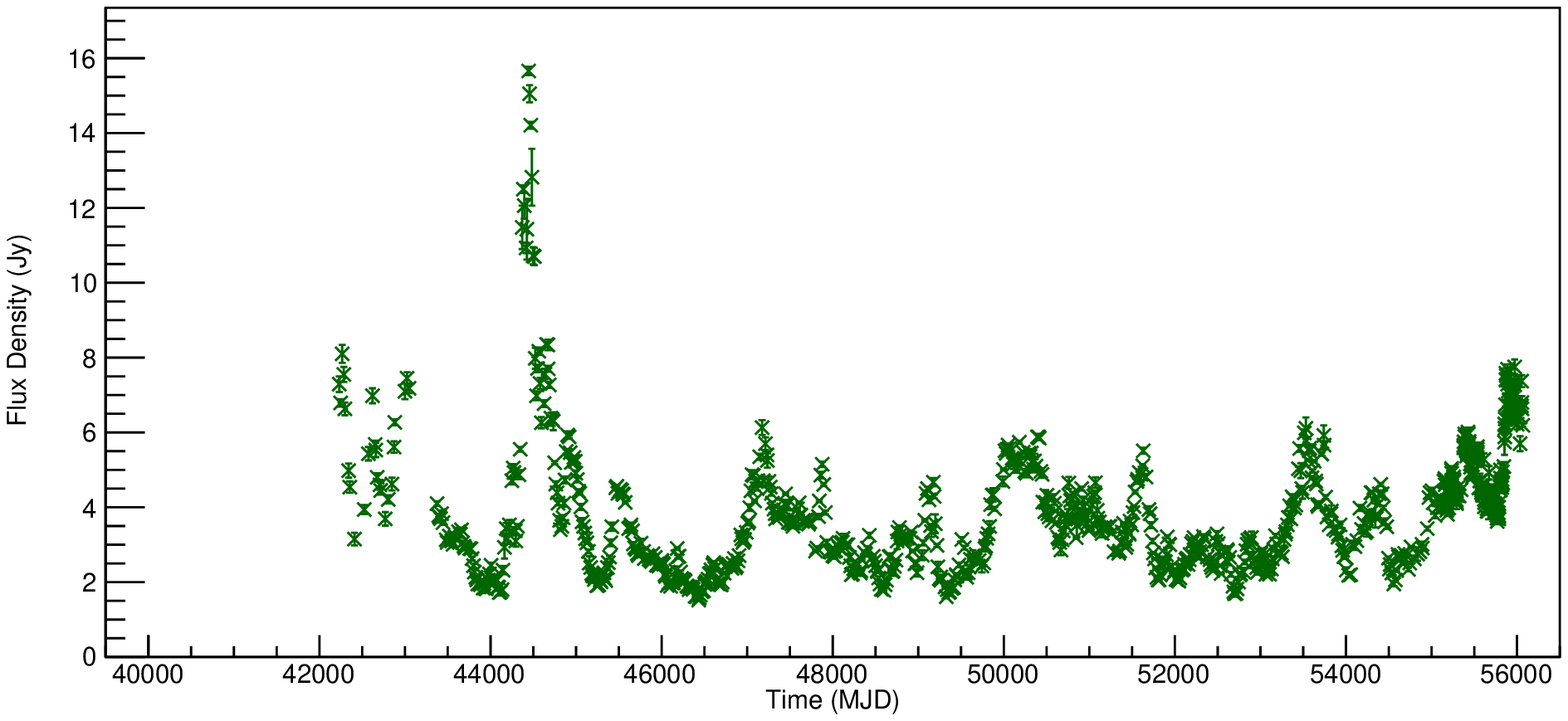}%
\caption{\footnotesize Light curve of PKS~2200+420 radio source, in $14.5~\mathrm{GH\lowercase{z}}$, shaw the raw dataset.}
\label{fig:rawbllac145}
\end{figure}

The Fig.~\ref{fig:rawbllac048} and~\ref{fig:rawbllac145} show light curve of PKS~2200+420 radio source for $4.8~\mathrm{GH\lowercase{z}}$ and $14.5~\mathrm{GH\lowercase{z}}$ separately after preprocessing step.

\begin{figure}[!h]\centering
\includegraphics[width=1.0\linewidth]{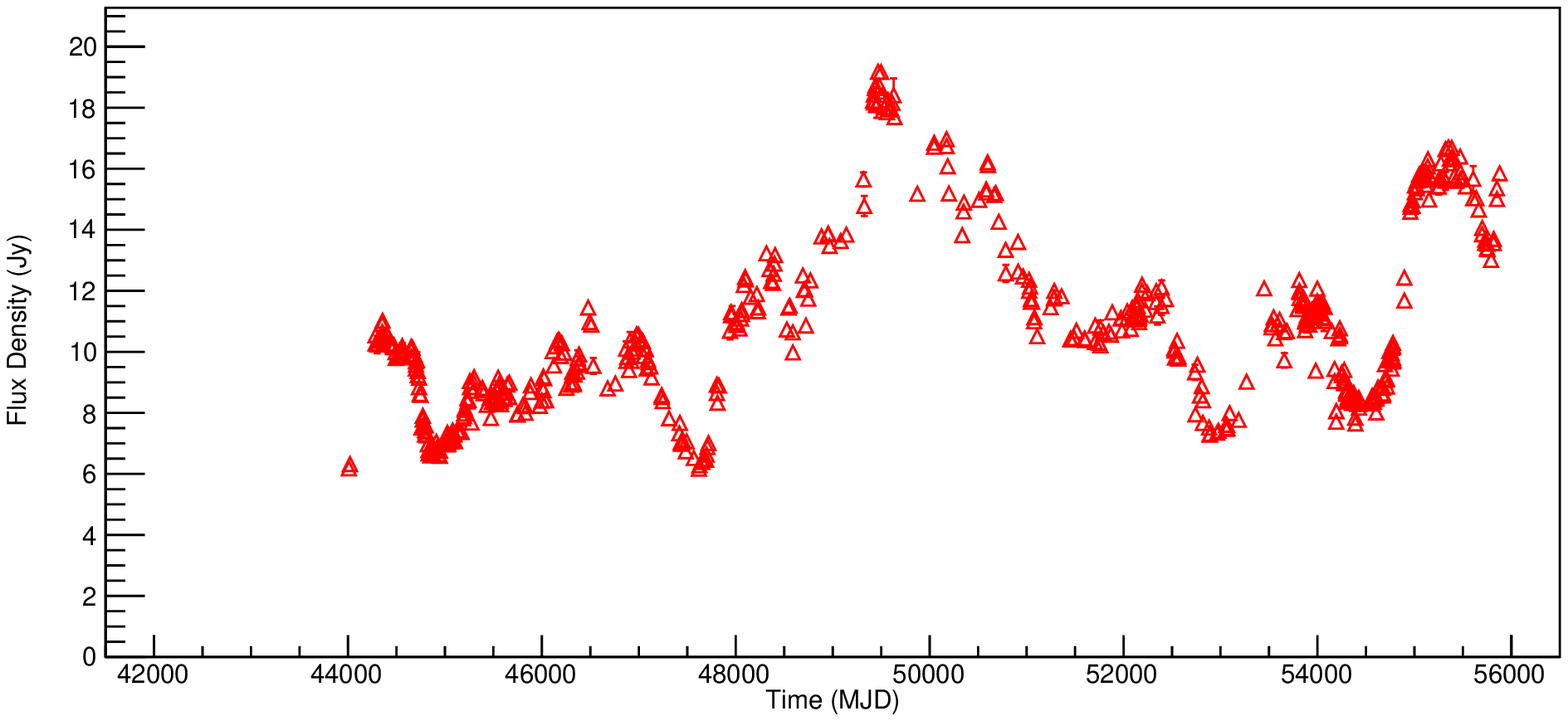}%
\caption{\footnotesize Light curve of PKS~1921-293 radio source in $4.8~\mathrm{GH\lowercase{z}}$, saw the raw dataset.}
\label{fig:rawov236048}
\end{figure}

\begin{figure}[!h]\centering
\includegraphics[width=1.0\linewidth]{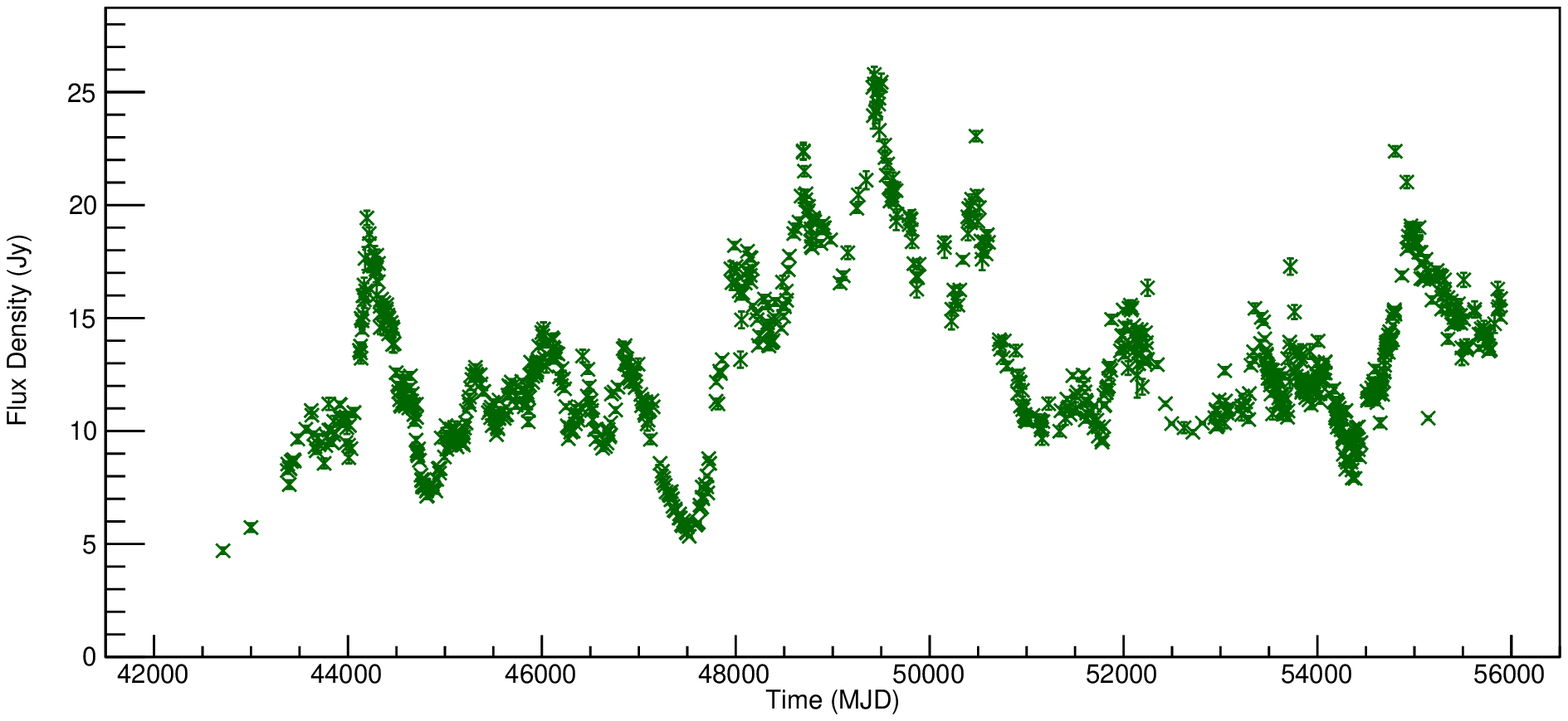}%
\caption{\footnotesize Light curve of PKS~1921-293 radio source in $14.5~\mathrm{GH\lowercase{z}}$, saw the raw dataset.}
\label{fig:rawov236145}
\end{figure}

The Fig.~\ref{fig:rawov236048} and~\ref{fig:rawov236145} show light curve of PKS~1921-293 radio source for $4.8~\mathrm{GH\lowercase{z}}$ and $14.5~\mathrm{GH\lowercase{z}}$ separately after preprocessing step.

\section{Time series regularized for objects PKS~2200+420 and PKS~1921-293}\label{sec:appRegular}

\begin{figure}[!h]\centering
\includegraphics[width=1.0\linewidth]{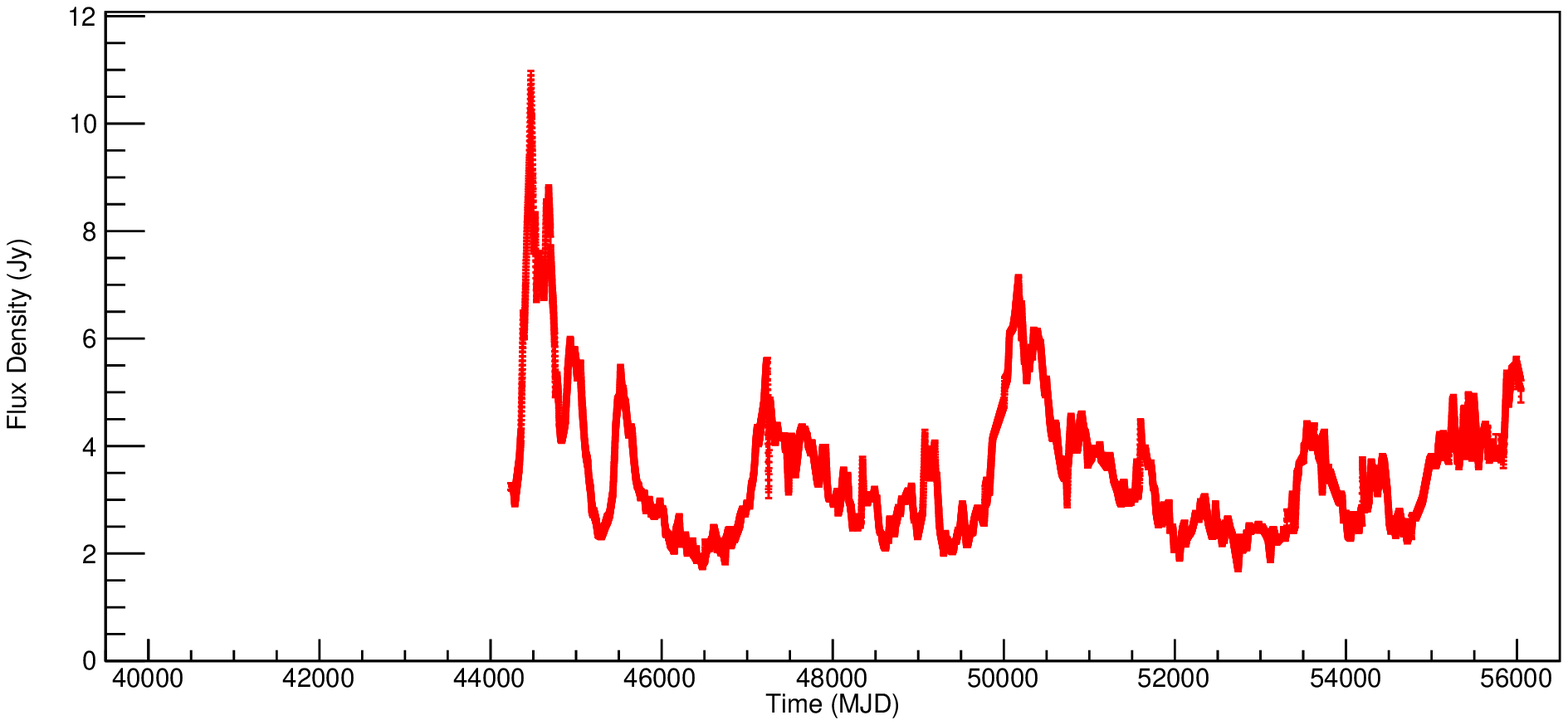}%
\caption{Light curve of PKS~2200+420 (BL~Lac) radio source, in $4.8~\mathrm{GH\lowercase{z}}$, shaw the space-time series regularized.}
\label{fig:regbllac048}
\end{figure}

\begin{figure}[!h]\centering
\includegraphics[width=1.0\linewidth]{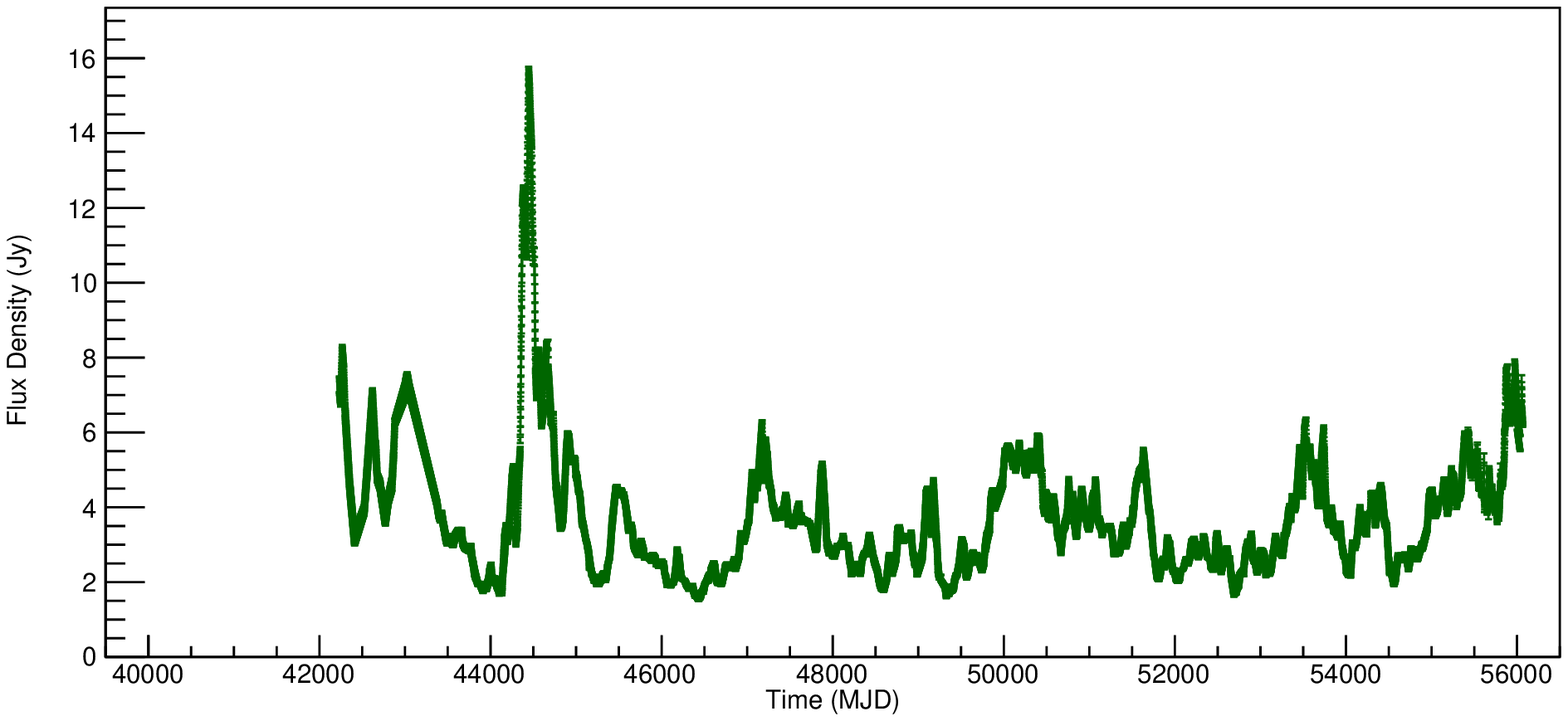}%
\caption{Light curve of PKS~2200+420 (BL~Lac) radio source, in $14.5~\mathrm{GH\lowercase{z}}$, shaw the space-time series regularized.}
\label{fig:regbllac145}
\end{figure}

The Fig.~\ref{fig:regbllac048} and~\ref{fig:regbllac145} show light curve of PKS~2200+420 radio source for $4.8~\mathrm{GH\lowercase{z}}$ and $14.5~\mathrm{GH\lowercase{z}}$ after regularization of time series step of method.

\begin{figure}[!h]\centering
\includegraphics[width=1.0\linewidth]{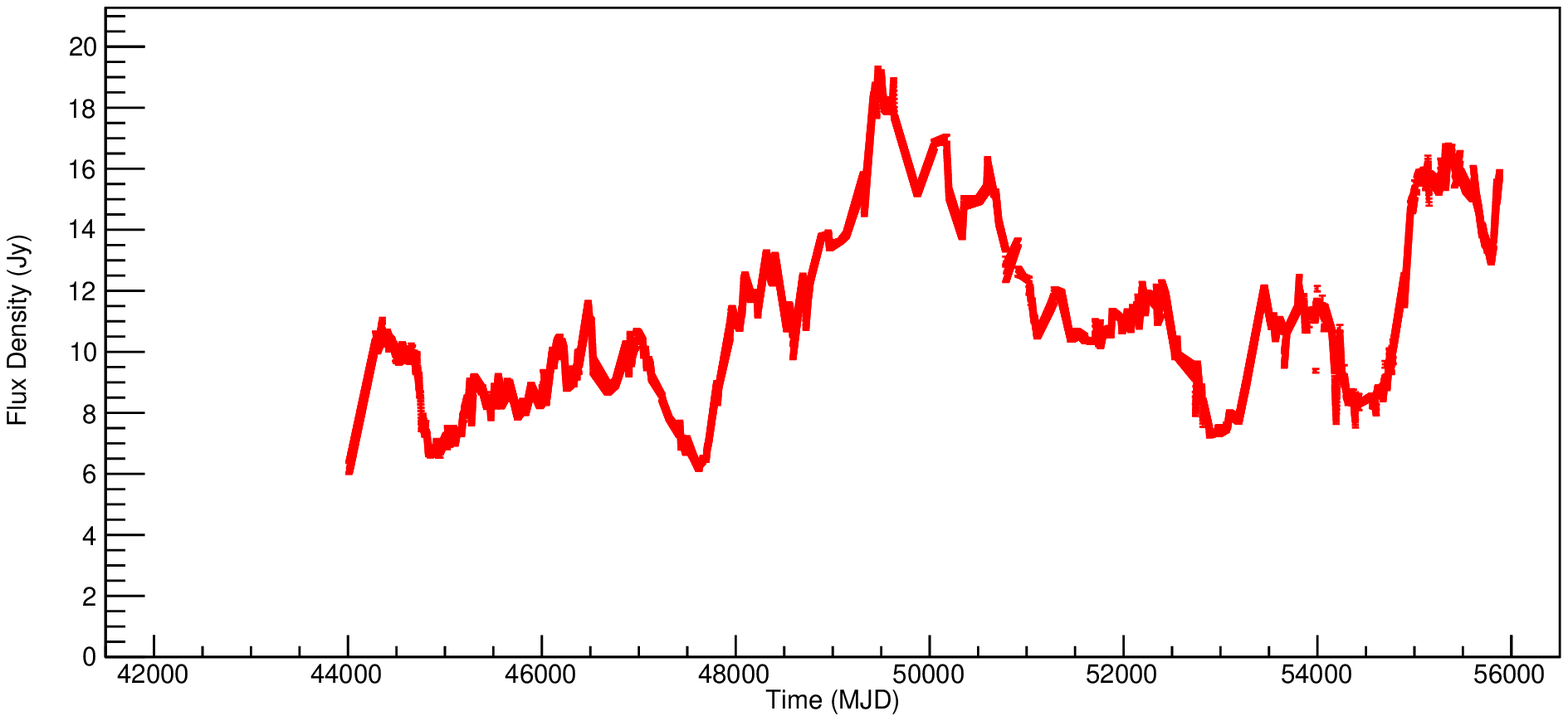}%
\caption{Light curve of PKS~1921-293 (OV~236) radio source in $4.8~\mathrm{GH\lowercase{z}}$, saw the space-time series regularized.}
\label{fig:regov236048}
\end{figure}

\begin{figure}[!h]\centering
\includegraphics[width=1.0\linewidth]{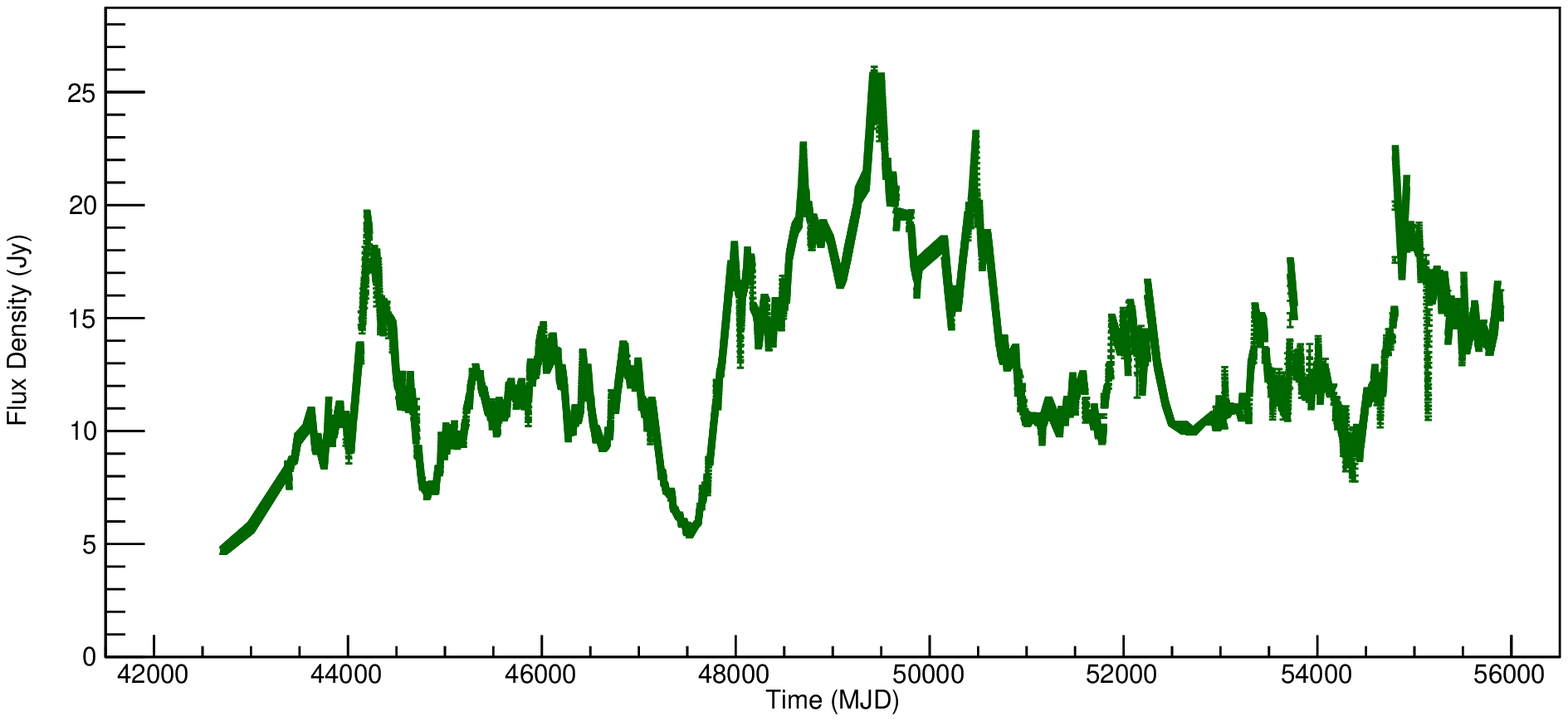}%
\caption{Light curve of PKS~1921-293 (OV~236) radio source in $14.5~\mathrm{GH\lowercase{z}}$, saw the space-time series regularized.}
\label{fig:regov236145}
\end{figure}

The Fig.~\ref{fig:regov236048} and~\ref{fig:regov236145} show light curve of PKS~1921-293 radio source for $4.8~\mathrm{GH\lowercase{z}}$ and $14.5~\mathrm{GH\lowercase{z}}$ after regularization of time series step of method.

\section{Finding outbursts in light curves of objects PKS~2200+420 and PKS~1921-293}\label{sec:appOutBurst}

\begin{figure}[!h]\centering
\includegraphics[width=1.0\linewidth]{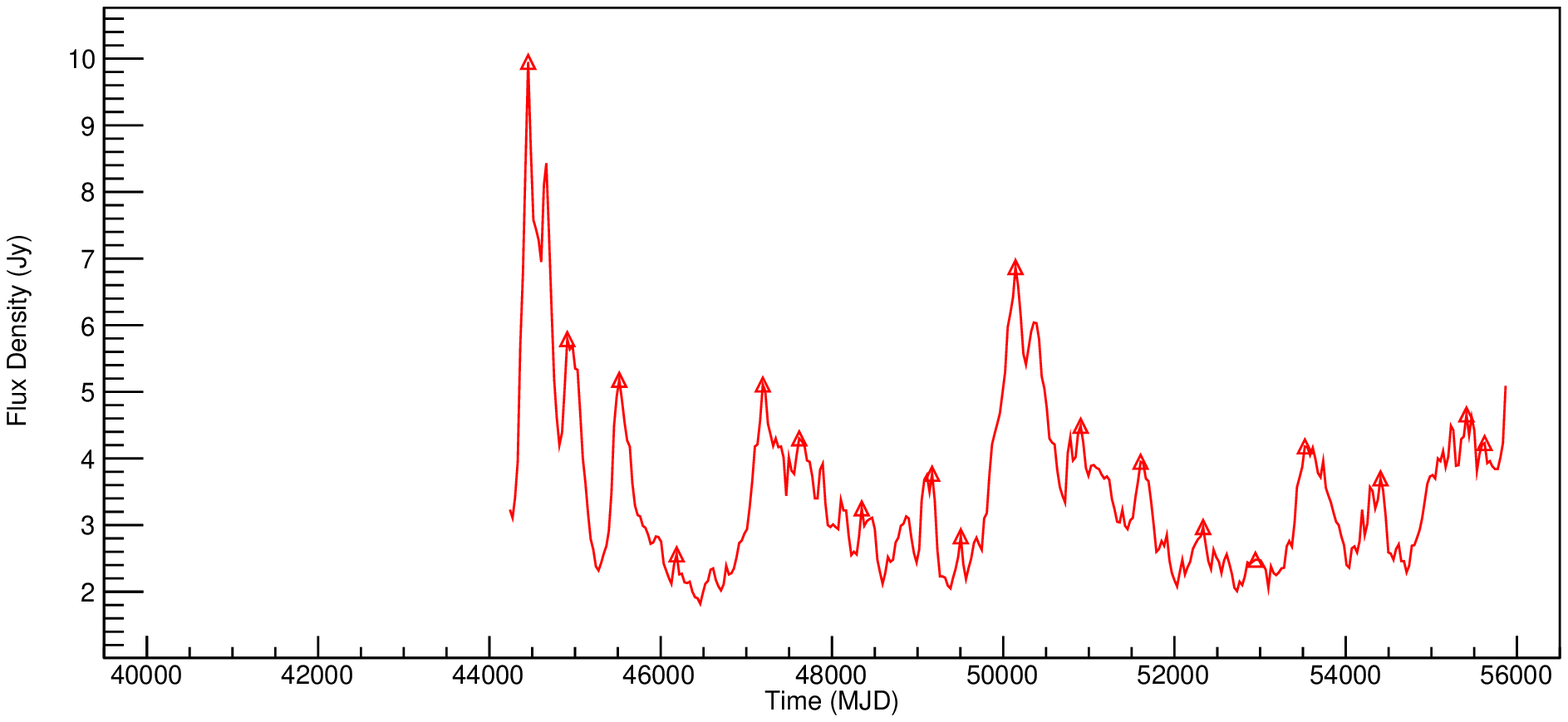}%
\caption{Light curve of PKS~2200+420 (BL~Lac) radio source, in $4.8~\mathrm{GH\lowercase{z}}$, shaw the detected outbursts to found periodicities.}
\label{fig:peakbllac048}
\end{figure}

\begin{figure}[!h]\centering
\includegraphics[width=1.0\linewidth]{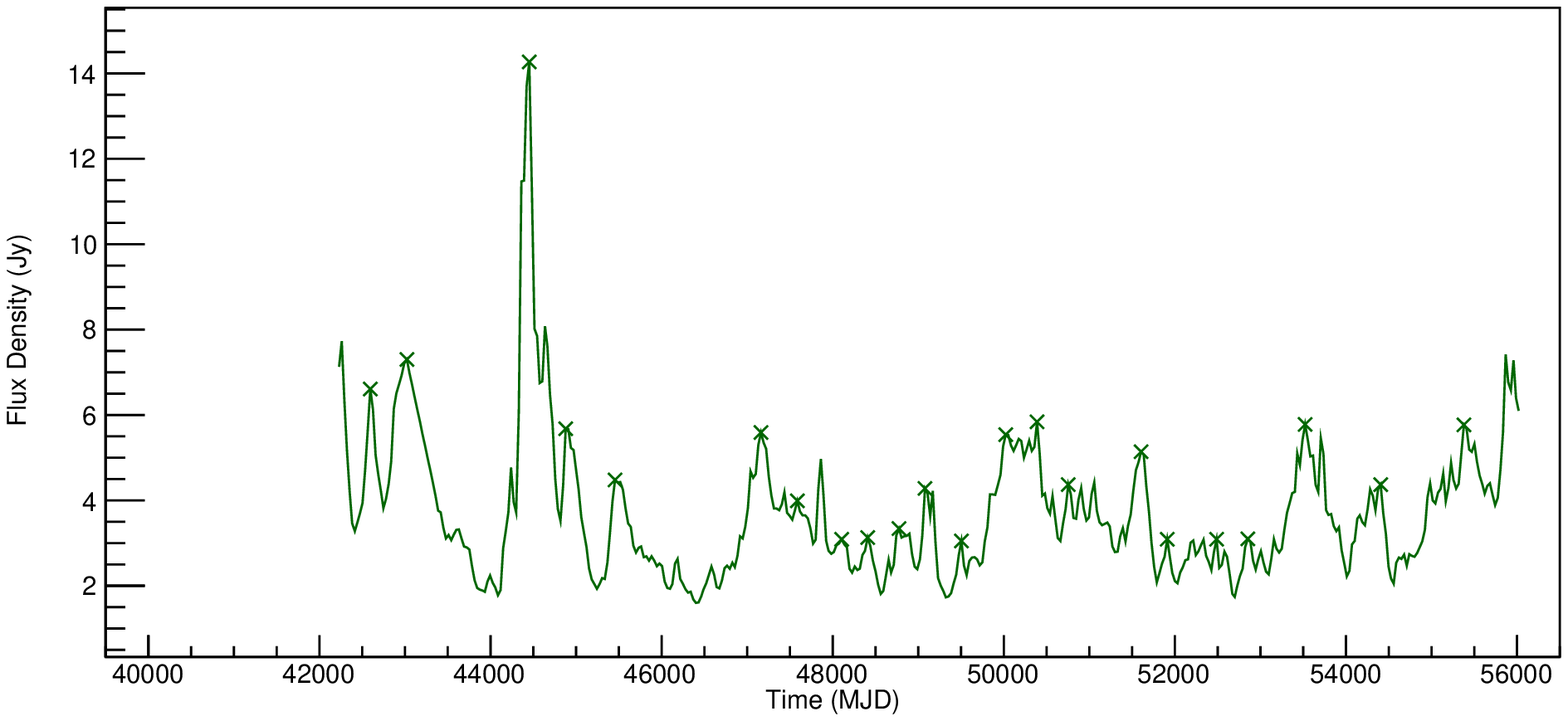}%
\caption{Light curve of PKS~2200+420 (BL~Lac) radio source, in $14.5~\mathrm{GH\lowercase{z}}$, shaw the detected outbursts to found periodicities.}
\label{fig:peakbllac145}
\end{figure}

At the end of the methodological process of looking for explosions, the outburst candidates detected by the algorithm for PKS~2200+420 (BL~Lac) radio source, in frequencies $4.8~\mathrm{GH\lowercase{z}}$ and $14.5~\mathrm{GH\lowercase{z}}$, were plotted in graph, as shown in the Figs.~\ref{fig:peakbllac048} and~\ref{fig:peakbllac145}.

\begin{figure}[!h]\centering
\includegraphics[width=1.0\linewidth]{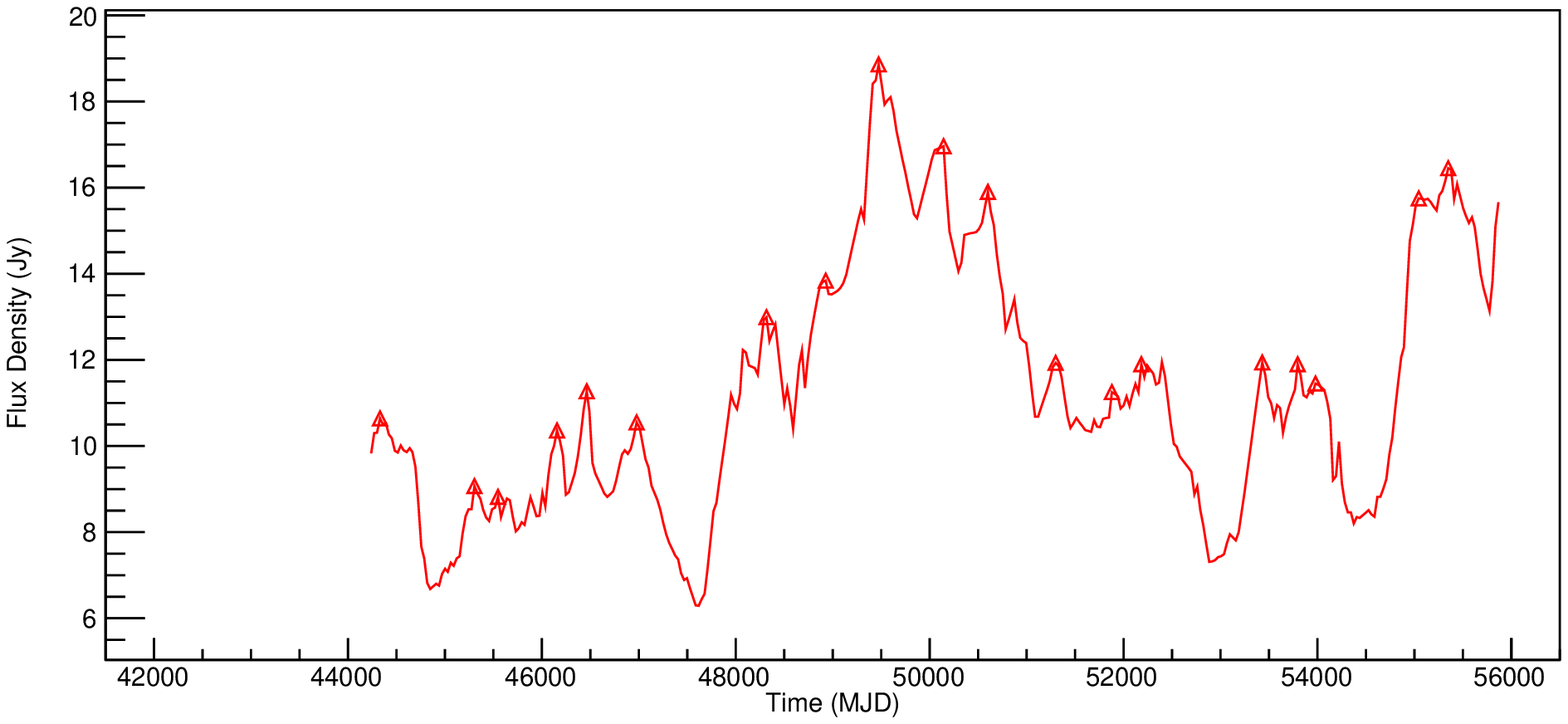}%
\caption{Light curve of PKS~1921-293 (OV~236) radio source in $4.8~\mathrm{GH\lowercase{z}}$, saw the detected outbursts to found periodicities.}
\label{fig:peakov236048}
\end{figure}

\begin{figure}[!h]\centering
\includegraphics[width=1.0\linewidth]{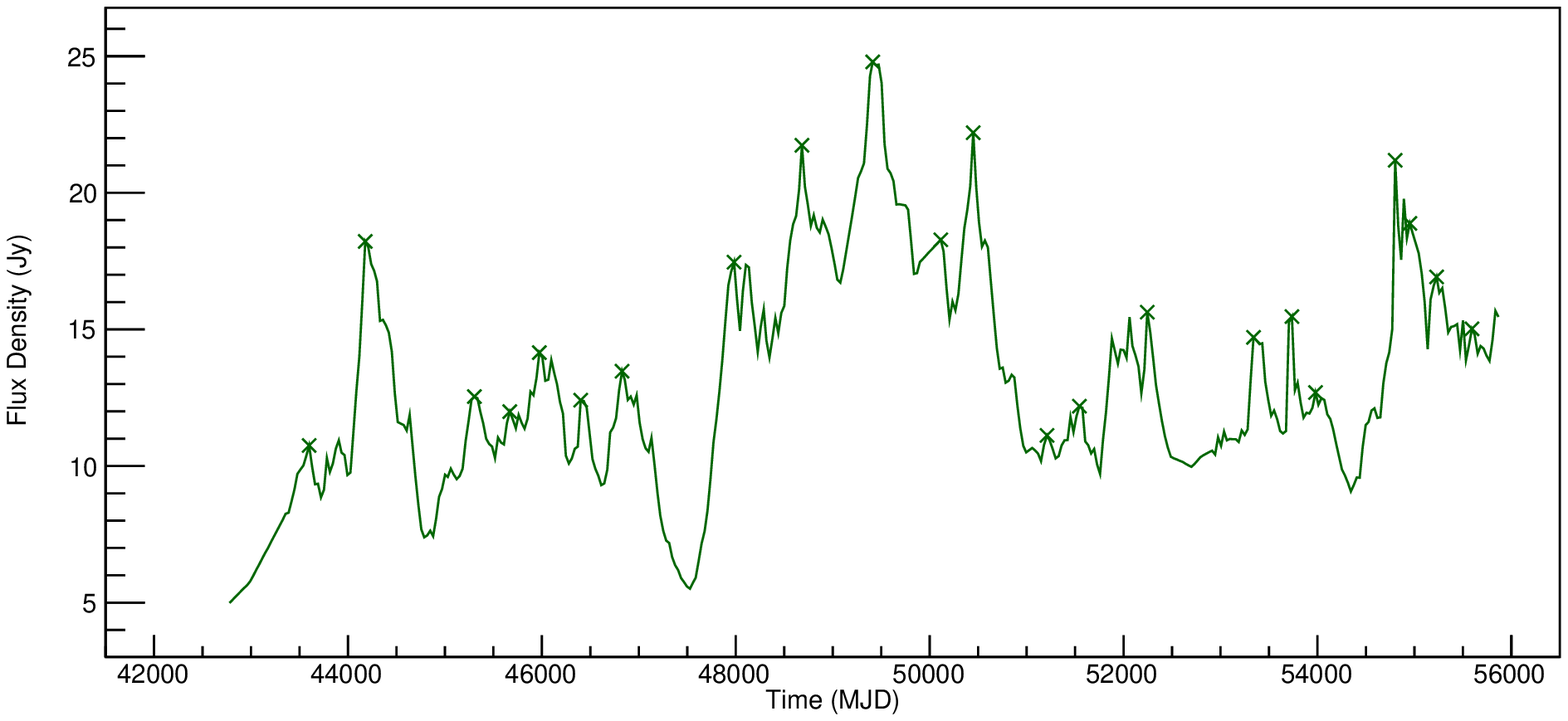}%
\caption{Light curve of PKS~1921-293 (OV~236) radio source in $14.5~\mathrm{GH\lowercase{z}}$, saw the detected outbursts to found periodicities.}
\label{fig:peakov236145}
\end{figure}

The same procedure was done for PKS~1921-293 (OV~236) radio source, at frequencies  $4.8~\mathrm{GH\lowercase{z}}$ and $14.5~\mathrm{GH\lowercase{z}}$, as shown in Figs.~\ref{fig:peakov236048} and~\ref{fig:peakov236145}.
\end{appendices}


\begin{thebibliography}
\bibitem[Aarshay (2016)]{Aarshay2016} Aarshay, J. 2016, Complete guide to parameter tuning in XGBoost, access date: 6 mar. 2019.
\bibitem[Abay et al. (2018)]{AbayBoyce2018} Abay, R., Boyce, R., Brown, M., \& Gehly, S. 2018, in 42nd COSPAR Scientific Assembly, Vol. 42, PEDAS.1-16-18
\bibitem[Abraham et al. (1982)]{AbrahamKaufmannBotti1982} Abraham, Z., Kaufmann, P., \& Botti, L. C. L. 1982, \aj, 87, 532
\bibitem[Aller (1992)]{Aller1992} Aller, H. D. in , Observing At A Distance Proceedings Of A Workshop On Remote Observing, ed. D. T. Emerson R. G. Clowes (Singapore: World Scientific Publishing Company), 31-35, april 21-23, 1992 -- Tucson, Arizona, USA
\bibitem[Aller et al. (2017)]{Aller2017} Aller, H. D. \& Aller, M. 2011, in \baas, Vol. 43, American Astronomical Society Meeting Abstracts, 142.47
\bibitem[Aller \& Aller (2010)]{AllerAller2010} Aller, H. D. \& Aller, M. F. 2010, in \baas, Vol. 42, American Astronomical Society Meeting Abstracts, 378
\bibitem[Aller \& Aller (2011)]{AllerAller2011} Aller, M. F., Aller, H. D., \& Hughes, P. A. 2009, in \baas, Vol. 41, American Astronomical Society Meeting Abstracts, 331
\bibitem[Aller et al. (2009)]{AllerAllerHughes2009} Aller, M. F., Aller, H. D., \& Hughes, P. A. 2017, Galaxies, 5, 75
\bibitem[Antonucci (1993)]{Antonucci1993} Antonucci, R. R. J. 1993, Annual Review of \aap, 31, 473
\bibitem[Askar et al. (2019)]{Askar2019} Askar, A., Askar, A., Pasquato, M., \& Giersz, M. 2019, \mnras, 485, 5345
\bibitem[Bakoyannis (2019)]{Bakoyannis2019} Bakoyannis, G. 2019, arXiv e-prints
\bibitem[Beckmann \& Shrader (2012)]{BeckmannShrader2012} Beckmann, V. \& Shrader, C. 2012, Active Galactic Nuclei (Weinheim, Germany: Wiley-VCH)
\bibitem[Bethapudi \& Desai (2018)]{Bethapudi2018} Bethapudi, S. \& Desai, S. 2018, Astronomy and Computing, 23, 15
\bibitem[Botti (1983)]{Botti1983} Botti, L. C. L. 1983, NASA STI/Recon Technical Report N, 84
\bibitem[Botti (1990)]{Botti1990} Botti, L. C. L. 1990, phdthesis, Instituto de Pesquisas Espaciais, São José dos Campos (Brazil)
\bibitem[Botti (1994)]{BottiIAU1994} Botti, L. C. L. 1994 in , IAU Colloquium (Ishiguro, M. and Welch, J.), 50, 140
\bibitem[Botti \& Abraham (1987)]{BottiAbraham1987} Botti, L. C. L. \& Abraham, Z. 1987, \RMAA, 14, 97
\bibitem[Botti \& Abraham (1988)]{BottiAbraham1988} Botti, L. C. L. \& Abraham, Z. 1988, \aj, 96, 465
\bibitem[Brighton \& Mellish (2002)]{Brighton2002} Brighton, H. \& Mellish, C. 2002, Data Mining and Knowledge Discovery, 6, 153
\bibitem[Caceres et at. (2019)]{Caceres2019} Caceres, G. A.; Feigelson, E. D.; Babu, G. J.; Bahamonde, N.; Christen, A.; Bertin, K.; Meza, C. \& Cur\'e, M. 2019, \aj, 158, 57
\bibitem[Calderon \& Berlind (2019)]{Calderon2019} Calderon, V. F. \& Berlind, A. A. 2019, \mnras, 490, 2367
\bibitem[Carruba et al. (2020)]{Carruba2020} Carruba, V., Aljbaae, S., Domingos, R. C., Lucchini, A., \& Furlaneto, P. 2020, \mnras, 496, 540
\bibitem[Chawla et al. (2002)]{Chawla2002} Chawla, N. V., Bowyer, K. W., Hall, L. O., \& Kegelmeyer, W. P. 2002, Journal Of Artificial Intelligence Research, 16, 321
\bibitem[Chen \& Guestrin (2016)]{Chen2016} Chen, T. \& Guestrin, C. 2016, in Proceedings of the 22Nd ACM SIGKDD International Conference on Knowledge Discovery and Data Mining, KDD -16 (New York, NY, USA: ACM Press), 785-794
\bibitem[Chong \& Yang (2019)]{ChongYang2019} Chong, K. \& Yang, A. 2019, EPJ Web of Conferences, 206, 9006
\bibitem[Ciaramella et al. (2004)]{Ciaramella2004} Ciaramella, A., Bongardo, C., Aller, H. D., Aller, M. F., De Zotti, G., Lahteenmaki, A., Longo, G., Milano, L., Tagliaferri, R., Terasranta, H., Tornikoski, M., \& Urpo, S. 2004, \aap, 419, 485
\bibitem[Cincotta et al. (1995)]{Cincotta1995} Cincotta, P. M., Méndez, M., \& Nunez, J. 1995, \apj, 449, 231
\bibitem[del Barrio et al. (2019)]{delBarrio2019} del Barrio, E., Inouzhe, H., \& Matrán, C. 2019, arXiv e-prints
\bibitem[Fan et al. (2007)]{FanLiu2007} Fan, J. H., Liu, Y., Yuan, Y. H., Hua, T. X., Wang, H. G., Wang, Y. X., Yang, J. H., Gupta, A. C., Li, J., Zhou, J. L., Xu, S. X., \& Chen, J. L. 2007, \aap, 462, 547
\bibitem[Gastaldi (2016)]{Gastaldi2016} Gastaldi, M. R. 2016, PhD thesis, Programa de Pós-Graduação em Ciências e Aplicações Geoespaciais da Universidade Presbiteriana Mackenzie, São Paulo (Brazil)
\bibitem[Hinkel et al. (2020)]{Hinkel2020} Hinkel, N., Unterborn, C., Kane, S., \& Somers, G. 2020, in American Astronomical Society Meeting Abstracts, 386.02
\bibitem[Hosenie et al. (2020)]{HosenieLyon2020} Hosenie, Z., Lyon, R., Stappers, B., Mootoovaloo, A., \& McBride, V. 2020, \mnras, 493, 6050
\bibitem[Huang, Yu-Pei, Yen, Meng-Feng(2019)]{HuangYen2019} Huang, Yu-Pei, Yen, Meng-Feng 2019, Applied Soft Computing, 83, 105663
\bibitem[Jin et al. (2019)]{JinZhang2019} Jin, X., Zhang, Y., Zhang, J., Zhao, Y., Wu, X.-b., \& Fan, D. 2019, \mnras, 485, 4539
\bibitem[Kelly et al. (2003)]{KellyHughes2003} Kelly, B. C., Hughes, P. A., Aller, H. D., \& Aller, M. F. 2003, \apj, 591, 695
\bibitem[Lam \& Kipping (2018)]{LamKipping2018} Lam, C. \& Kipping, D. 2018, \mnras, 476, 5692
\bibitem[LeCun et al. (2015)]{LeCunBengioHinton2015} LeCun, Y., Bengio, Y., \& Hinton, G. E. 2015, \nat, 521, 436
\bibitem[Li et al. (2019)]{Li2019} Li, Chao, L., Zhang, W.-h., \& Lin, J.M. 2019, Chinese \aap, 43, 539
\bibitem[Li et al. (2020)]{Li2020} Li, C., Zhang, W. H., \& Lin, J. M. 2020, Acta Astronomica Sinica, 61, 21
\bibitem[Lin et al. (2020)]{LinLiLuo2020} Lin, H., Li, X., \& Luo, Z. 2020, \mnras, 493, 1842
\bibitem[Liu et al. (2019)]{LiuRyley2019} Liu, R. H., Hill, R., Scott, D., Almaini, O., An, F., Gubbels, C., Hsu, L.-T., Lin, L., Smail, I., \& Stach, S. 2019, \mnras, 489, 1770
\bibitem[Marsaglia et al. (2003)]{Marsaglia2003} Marsaglia, G., Tsang, W. W., \& Wang, J. 2003, Journal of Statistical Software, Articles, 8, 1
\bibitem[Matthews \& Sandage (1963)]{MatthewsSandage1963} Matthews, T. A. \& Sandage, A. R. 1963, \apj, 138, 30
\bibitem[Menou (2019)]{Menou2019} Menou, K. 2019, \mnras, 489, 4802
\bibitem[Mitchell \& Frank (2017)]{MitchellFrank2017} Mitchell, R. \& Frank, E. 2017, PeerJ Computer Science, 3, e127
\bibitem[Pashchenko et al. (2017)]{Pashchenko2017} Pashchenko, I. N., Sokolovsky, K. V., \& Gavras, P. 2017, \mnras, 475, 2326
\bibitem[Plavin et al. (2019)]{PlavinKovalev2019} Plavin, A. V., Kovalev, Y. Y., Pushkarev, A. B., \& Lobanov, A. P. 2019, \mnras, 485, 1822
\bibitem[Rasheed et al. (2011)]{RasheedAlshalalfaAlhajj2011} Rasheed, F., Alshalalfa, M., \& Alhajj, R. 2011, IEEE Transactions on Knowledge and Data Engineering, 23, 79
\bibitem[van Roestel et al. (2018)]{Roestel2018} van Roestel, J., Kupfer, T., Ruiz-Carmona, R., Groot, P. J., Prince, T. A., Burdge, K., Laher, R., Shupe, D. L., \& Bellm, E. 2018, \mnras, 475, 2560
\bibitem[Sadhanala et al. (2019)]{Sadhanala2019} Sadhanala, V., Wang, Y.-X., Ramdas, A., \& Tibshirani, R. J. 2019, arXiv e-prints
\bibitem[Saha et al. (2018)]{Saha2018} Saha, S., Basak, S., Safonova, M., Bora, K., Agrawal, S., Sarkar, P., \& Murthy, J. 2018, Astronomy and Computing, 23, 141
\bibitem[Santos (2007)]{Auta2007} Santos, M. A. d. 2007, mathesis, Mackenzie Presbyterian University, São Paulo
\bibitem[Schmidt (1963)]{Schmidt1963} Schmidt, M. 1963, \nat, 197, 1040
\bibitem[Shu et al. (2019)]{ShuKoposov2019} Shu, Y., Koposov, S. E., Evans, N. W., Belokurov, V., McMahon, R. G., Auger, M. W., \& Lemon, C. A. 2019, \mnras, 489, 4741
\bibitem[Smirnov \& Markov (2017)]{Smirnov2017} Smirnov, E. A. \& Markov, A. B. 2017, \mnras, 469, 2024
\bibitem[Soldi et al. (2008)]{Soldi2008} Soldi, S., Türler, M., Paltani, S., Aller, H. D., Aller, M. F., Burki, G., Chernyakova, M., Lähteenmäki, A., McHardy, I. M., Robson, E. I., Staubert, R., Tornikoski, M., Walter, R., \& Courvoisier, T. J.-L. 2008, \aap, 486, 411
\bibitem[Tamayo et al. (2020)]{Tamayo2020} Tamayo, D., Cranmer, M., Hadden, S., Rein, H., Battaglia, P., Obertas, A., Armitage, P. J., Ho, S., Spergel, D. N., Gilbertson, C., Hussain, N., Silburt, A., Jontof-Hutter, D., \& Menou, K. 2020, Proceedings of the National Academy of Sciences
\bibitem[Tornikoski et al. (1996)]{Tornikoski1996} Tornikoski, M., Valtaoja, E., Teraesranta, H., Karlamaa, K., Lainela, M., Nilsson, K., Kotilainen, J., Laine, S., Laehteenmaeki, A., Knee, L. B. G., \& Botti, L. C. L. 1996, \aaps, 116, 157
\bibitem[Tsizh et al. (2020)]{Tsizh2020} Tsizh, M., Novosyadlyj, B., Holovatch, Y., \& Libeskind, N. I. 2020, \mnras, 495, 1311
\bibitem[Urry \& Padovani (1995)]{UrryPadovani1995} Urry, C. M. \& Padovani, P. 1995, Publications of the Astronomical Society of the Pacific, 107, 803
\bibitem[Véron-Cetty \& Véron (2010)]{VeronCetty2010} Véron-Cetty, M.-P. \& Véron, P. 2010, \aap, 518, A10
\bibitem[Villata et al. (2004)]{Villata2004} Villata, M., Raiteri, C. M., Aller, H. D., Aller, M. F., Teräsranta, H., Koivula, P., Wiren, S., Kurtanidze, O. M., Nikolashvili, M. G., Ibrahimov, M. A., Papadakis, I. E., Tosti, G., Hroch, F., Takalo, L. O., Sillanpää, A., Hagen-Thorn, V. A., Larionov, V. M., Schwartz, R. D., Basler, J., Brown, L. F., \& Balonek, T. J. 2004, \aap, 424, 497
\bibitem[Vitoriano \& Botti (2018)]{Vitoriano2018} Vitoriano, R. P. \& Botti, L. C. L. 2018, \apj, 854, 59
\bibitem[Wang et al. (2019)]{WangPan2019} Wang, Y., Pan, Z., Zheng, J., Qian, L., \& Li, M. 2019, \apss, 364, 139
\bibitem[Witten et al. (2016)]{Witten2016} Witten, I. H., Frank, E., Hall, M. A., \& Pal, C. J. 2016, Data Mining: Practical Machine Learning Tools and Techniques, The Morgan Kaufmann Series in Data Management Systems (Cambridge, MA: Elsevier Science)
\bibitem[Xu (2018)]{Kaggle2018} Xu, B. 2018, Higgs Boson Machine Learning Challenge, access date: 28 ago. 2018
\bibitem[Yi et al. (2019)]{YiZesheng2019} Yi, Z., Chen, Z., Pan, J., Yue, L., Lu, Y., Li, J., \& Luo, A. L. 2019, \apj, 887, 241
\bibitem[Yuan (2011)]{Yuan2011} Yuan, Y. 2011, Journal of Astrophysics and Astronomy, 32, 43
\bibitem[Zhang et al. (2018)]{ZhangQianMaoHuangHuangSi2018} Zhang, D., Qian, L., Mao, B., Huang, C., Huang, B., Si, Y. 2018, IEEE Access, vol. 6, pp. 21020-21031
\end{thebibliography}
\end{document}